\documentclass [smallextended,natbib]{svjour3}
\usepackage{amsmath}
\usepackage{graphicx,psfrag,epsf}
\usepackage{enumerate}
\usepackage{natbib}

\usepackage{multicol}
\usepackage{graphicx}
\usepackage{epstopdf}
\usepackage{amssymb}
\usepackage{amsmath}
\usepackage{caption}
\usepackage{mathtools}
\usepackage{caption}
\usepackage{subcaption}
\usepackage{tabularx}

\usepackage[hyphens]{url}
\usepackage[hidelinks]{hyperref}
\hypersetup{breaklinks=true}
\DeclareMathOperator{\var}{var}
\DeclareMathOperator{\diag}{diag}
\DeclareMathOperator{\cov}{cov}

\def\makeheadbox{}

\def\figsize{0.75}

\begin{document}

\journalname{Computational Statistics}
\date{27 May 2018}
\title{Efficient convergence through adaptive learning in sequential Monte Carlo Expectation Maximization}
\titlerunning{Adaptive learning in sequential Expectation Maximization}
\author{Donna Henderson \and Gerton Lunter}
\institute{Wellcome Centre of Human Genetics, University of Oxford, Oxford OX3 7BN, UK. \\\email{gerton.lunter@well.ox.ac.uk}}

%

\maketitle

\begin{abstract}
Expectation maximization (EM) is a technique for estimating maxi\-mum-likelihood parameters
of a latent variable model given observed data
by alternating between taking expectations of sufficient statistics,
and maximizing the expected log likelihood.
For situations where sufficient statistics are intractable,
stochastic approximation EM (SAEM)
is often used, which uses Monte Carlo techniques to approximate
the expected log likelihood.  Two common implementations of
SAEM, Batch EM (BEM) and online EM (OEM), are parameterized by a
``learning rate'',
and their efficiency depend strongly on this parameter.
We propose an extension to the OEM algorithm, termed Introspective
Online Expectation Maximization (IOEM),
which removes the need for specifying this parameter by adapting
the learning rate according to trends in the parameter updates.
We show that our algorithm matches the efficiency of the optimal BEM
and OEM algorithms in multiple models, and that the efficiency of IOEM
can exceed that of BEM/OEM methods with optimal learning rates
when the model has many parameters.
A Python implementation is available at
https://github.com/luntergroup/IOEM.git.


\end{abstract}

\begin{acknowledgements}
This work was supported by Wellcome Trust grants 090532/Z/09/Z and 102423/Z/13/Z.
\end{acknowledgements}



\section{Introduction}
\label{intro}

Expectation Maximization (EM) is a widely used and general technique for estimating maximum likelihood parameters of a latent
variable model \citep{Dempster1977maximum}. We will be considering models with a sequential structure.
Elegant algorithms are available for special cases of sequential models, such as linear systems with Gaussian
noise \citep{shumway1982approach}, and finite-state hidden Markov models \citep{baum1972equality}.
Here we focus on inference in complex models that do not admit analytic solutions, for which
sequential Monte Carlo (SMC) methods are widely used to approximate the expectation in the E-step.
Generally, the use of Monte Carlo methods in the context of EM is known as stochastic approximation EM
(SAEM; \citealt{delyon99})
and this class of methods is favored in practice over gradient-based approaches due to their relative stability and 
computational efficiency when estimating high dimensional parameters \citep{chitralekha2010comparison,kantas2009overview}.

Convergence of EM methods can nevertheless be slow for complex models and/or with large data volumes.  
Several authors have proposed acceleration techniques \citep{jamshidian93,lange95,varadhan08}, but these require
that the E-step is analytically tractable.  For SAEM standard recursive EM methods are used instead, the two
most popular being batch EM (BEM) and online EM (OEM).
Both methods require the user to specify a tuning parameter, and in both cases the performance of the algorithm
is strongly dependent on the chosen parameter.  For instance, for BEM, very large batch sizes lead to inaccurate
estimates because of slow convergence, whereas very small batch sizes lead to imprecise estimates due to
the inherent stochasticity of the model within a small batch of observations.
The optimal batch size in BEM, or equivalently the optimal learning rate in OEM,
depends on the particularities of the model.

While the relative merits of these and other methods for parameter estimation have been studied in detail
(see e.g.\ \citealt{kantas2009overview}), the problem of choosing optimal learning rates has received relatively little
attention. Here we introduce a novel algorithm, termed Introspective Online EM (IOEM), which removes the need
for setting the learning rate altogether by
estimating the optimal parameter-specific learning rate along with the parameters of interest.
This is particularly helpful when inferring parameters in a high dimensional
model, since the optimal tuning parameter may differ between parameters.  Broadly, IOEM works by
estimating both the precision and the accuracy of parameters in an online manner through weighted
linear regression, and uses these estimates to
tune the learning rate so as to improve both simultaneously.

The outline of this paper is as follows. Sect.~\ref{sec:sar} uses a one-parameter
autoregressive state-space model to introduce BEM, OEM, and a simplified version of IOEM.
Sect.~\ref{sec:ar} considers the full 3-parameter autogressive model, which requires the complete IOEM algorithm. 
Sect.~\ref{sec:2dar} considers a 2-dimensional autoregressive model to show the benefit of the proposed algorithm when inferring many parameters.
Finally, Sect.~\ref{sec:sv} demonstrates desirable performance in the stochastic volatility model,
an important case as it is nonlinear and hence more similar to applications of SAEM.

\section{EM for a Simplified Autoregressive Model}
\label{sec:sar}

Here we review SMC, BEM, OEM, and present the IOEM algorithm with a simple model.
This illustrates the main
concepts behind IOEM before delving into details in Sect. \ref{sec:ar}.

We consider a simple autoregressive model with one unknown parameter.
We observe the sequence of random variables $Y_{1:t}:=\{Y_k\}_{k=1,...,t}$ which depends on the
unobserved sequence $X_{1:t}:=\{X_k\}_{k=1,...,t}$, as follows:
\begin{align}
X_t&=a X_{t-1}+ \sigma_w W_t, \nonumber \\
Y_t&=X_t + \sigma_v V_t,
\label{ar}
\end{align}
where $W_t$ and $V_t$ are i.i.d.\ standard normal variates, $a=0.95$ and $\sigma_w^2=1$ are known
parameters, and $\sigma_v^2$ is unknown.
Under this model, we have the following
transition and emission densities:
\begin{align*}
f(x_t|x_{t-1}) &= (2 \pi \sigma_w^2)^{-1/2} \exp\Big\{-\frac{(x_t - a x_{t-1})^2}{2 \sigma_w^2}\Big\}, \\
g(y_t|x_t) &= (2 \pi \sigma_v^2)^{-1/2} \exp\Big\{-\frac{(y_t- x_t)^2}{2 \sigma_v^2}\Big\}.
\end{align*}
We have chosen $\sigma_v^2$ as the unknown parameter as it is the most straightforward to estimate,
allowing us to introduce the idea of IOEM without certain complications which we address in Sect.~\ref{sec:ar}.
As $f$ and $g$ are members of the exponential family of distributions, the M step of EM can be
done using sufficient statistics, and so the E step amounts to the expectation of the sufficient statistics.
In this model, the parameter $\sigma_v^2$ has the sufficient statistic
\begin{equation} \label{S4}
    S_t = \mathbb{E}_{X_{1:t} | Y_{1:t}, \theta }\left[\frac{1}{t} \sum_{k=1}^{t}(Y_k-X_k)^2 \right].
\end{equation}
The estimate of $\sigma_v^2$ is obtained by setting $\hat{\sigma}_{v,t}^2=\hat{S}_t$.
More generally, for an unknown parameter $\theta$, $\hat{\theta}_t=\Lambda(\hat{S}_t)$
where $\Lambda$ is a known function mapping sufficient statistics to parameter estimates.

To estimate $S_t$, we use sequential Monte Carlo (SMC) to simulate
particles $X^{(i)}_{1:t}$ and their associated weights $w(X^{(i)}_{1:t})$, $i=1,\ldots,N$, so that
\begin{equation}
  \sum_{i=1}^N w(X^{(i)}_{1:t}) \delta_{X^{(i)}_{1:t}}
  \label{approxdistrib}
\end{equation}
approximates the distribution $p(X_{1:t}|Y_{1:t},\theta)$.
The standard MCEM approximation of $p(X_{1:t}|Y_{1:t},\hat{\theta})$ would require storage of all
observations $Y_{1:t}$, the simulation of $X^{(i)}_{1:t}$ each time $\hat{\theta}$ is updated,
and ideally an increasing Monte Carlo sample size as the parameter estimates near convergence.
To avoid this, we employ SAEM which effectively averages over previous parameter estimates as
an alternative to generating a new Monte Carlo sample every time an estimate is updated, and hence
is more suitable to online inference. This method as proposed in \cite{cappe2009line} approximates
the expectation in \eqref{S4} recursively.

The outline of the SMC with EM algorithm we consider in this paper is as follows:

\vspace{4mm}
\noindent \hrulefill \\
\noindent \textbf{Algorithm 1} Sequential Importance Resampling (bootstrap filter) \\
\noindent \null \hrulefill \\
For time $t \geq 1$:
\begin{enumerate}
\item
$\text{For } i=1, \ldots ,N:$ \\ $ \text{ Sample } X_t^{(i)} \sim
\begin{cases}
\mu(\cdot|\hat{\theta}_{0}), & \text{if } t=1 \\
f( \cdot | X_{t-1}^{(i)}, \hat{\theta}_{t-1}), & \text{if } t\geq2\\
\end{cases}
$
\item Compute normalized weights satisfying \\
$w_t(X_{1:t}^{(i)}) \propto w_{t-1}(X_{1:t-1}^{(i)}) \cdot g(Y_t|X_t^{(i)},\hat{\theta}_{t-1})$
\item Update $\hat{\theta}_{t-1}$ to $\hat{ \theta}_{t}$ using chosen EM method
\item Resample particles if $ ESS<\frac{N}{2}$
\end{enumerate}
\hrulefill
\vspace{4mm}

\noindent Here $\mu(\cdot | \hat{ \theta}_{0})$ is the initial distribution for $X_1$,
$ESS$ is the effective sample size defined as
$[\sum_{i=1}^N w_t(X_{1:t}^{(i)})^{-2}]^{-1}$, 
$w_0(\cdot)=1/N$, 
and
$X_{t}^{(i)}$ is shorthand for the $t^\text{th}$ coordinate of $X_{1:t}^{(i)}$.
In models with multiple unknown parameters, each parameter is updated in step 3 of the algorithm,
however we will refer only to a single parameter $\theta$ to keep the notation simple.

Throughout this paper we follow common practice in using the fixed-lag technique in order
to reduce the mean square error between $S_t$ and $\hat{S}_t$ \citep{cappe2005use,cappe2007overview}.
In particular, we choose a lag $\Delta > 0$
and then at time $t$, using particles $X_{1:t}^{(i)}$ shaped by data $Y_{1:t}$,
estimate the $t-\Delta^{\text{th}}$ term of the summation in \eqref{S4}.
We will use $X_{1:t}^{(i)}(t-\Delta)$ to denote the $t-\Delta^\text{th}$ coordinate of the particle $X_{1:t}^{(i)}$,
but we will continue to write $X_{t}^{(i)}$ as a shorthand for $X_{1:t}^{(i)}(t)$.
(see Table \ref{table:notation} for an overview of notation used in this paper.)

The fixed-lag technique involves making the approximation
\begin{align} \label{approxS}
S_t &\approx \mathbb{E}_{X_{1:t}|Y_{1:t} , \theta }\left[\frac{1}{t-\Delta} \sum_{j=1}^{t-\Delta} s(Y_j,X_j) \right ] \nonumber \\ 
    &\approx \frac{1}{t-\Delta} \sum_{j=1}^{t-\Delta} \mathbb{E}_{X_{1:j+\Delta}| Y_{1:j+\Delta} , \hat{\theta} } \left[ \strut s(Y_j,X_j) \right] ,
\end{align}
where we assume that $S_t$ can be written as
\begin{align*}
S_t &= \mathbb{E}_{X_{1:t}|Y_{1:t},\theta} \sum_{j=1}^t s(Y_j,X_j)
\end{align*}
This allows $S_t$ to be updated in an online manner
by computing the componentwise sufficient statistics
\begin{align*}
\tilde{s}_t:&=  \mathbb{E}_{ X_{1:t}|Y_{1:t} , \theta } \left[ s( Y_{t-\Delta}, X_{1:t}(t-\Delta) ) \right] \\
&\approx \sum_i w_k(X_{1:t}^{(i)}) s(Y_{t-\Delta}, X_{1:t}^{(i)}(t-\Delta)),
\end{align*}
allowing $\hat{S}_t$ to be updated as
\begin{equation*}
\hat{S}_t = \gamma_t \cdot \tilde{s}_t  + (1-\gamma_t) \cdot \hat{S}_{t-1},
\end{equation*}
with some weight $\gamma_t$; in \eqref{approxS} $\gamma_t=1/(t-\Delta)$.
This approach is slightly different from that of \citep{cappe2005use}; see
Sect.\ \ref{sec:lag} for a discussion.

Choosing a large value of $\Delta$ allows SMC to use
many observations to improve the posterior distribution of $X_{t-\Delta}$.
However the cost of a large $\Delta$ is a loss in particle independence due to the
resampling procedure which increases the sample variance.
The optimal choice for $\Delta$ 
balances the opposing influences of the forgetting rate
of the model and the collapsing rate of the resampling process due
to the divergence between the proposal distribution and the posterior distribution.
For the examples in this paper we chose $\Delta = 20$ as recommended by \citet{cappe2005use},
which seems to be a reasonable choice for our models.

There are various other techniques to improve on this basic SMC method, including
improved resampling schemes \citep{douc2005comparison,olsson2008sequential,doucet2009tutorial,cappe2007overview},
and choosing better sampling distributions through lookahead strategies or resample-move procedures
\citep{pitt1999filtering,lin2013lookahead,doucet2009tutorial}, which are not discussed further here.
Instead, in the remainder of this paper, we focus on the process of updating the parameter estimates $\hat{ \theta}_{t}$.
The remainder of this section describes the options for step 3 of Algorithm 1.

\subsection{Batch Expectation Maximization}
\label{sec:bem}

Batch Expectation Maximization (BEM) processes the data in batches.  Within a batch
of size $b$, the parameter estimate stays constant ($\hat{ \theta}_{t}=\hat{ \theta}_{t-1}$) and the update
to the sufficient statistic
\begin{equation*}
\tilde{s}_{t}:=\sum_{i}w_t(X_{1:t}^{(i)}) \cdot (Y_{t-\Delta}-X_{1:t}^{(i)}(t-\Delta))^2,
\end{equation*}
is collected at each iteration $t$. At the end of the $m$th batch we have $t=mb$, at which time
\begin{equation*}
\hat{S}^{BEM}_{t}:=\frac{1}{b}\sum_{k=(m-1)b+1}^{mb}\tilde{s}_{k},
\end{equation*}
is our approximation of $S$, and $\hat{\sigma}^{2}_{v,t}:=\hat{S}^{BEM}_{t}$.

The batch size determines the convergence behavior of the estimates. For a fixed computational cost,
choosing $b$ too small will result in
noise-dominated estimates and low precision, whereas choosing $b$ too large will result
in precise but inaccurate estimates due to slow convergence.

\subsection{Online Expectation Maximization}
\label{sec:oem}

BEM only makes use of the collected evidence at the end of each batch, missing potential early opportunities for
improving parameter estimates.  OEM addresses this issue by updating the parameter estimate at every iteration.
The approximation of $S$ at time $t$ is a running average of
$\{\tilde{s}_{k}\}_{k=\Delta+1,...,t}$, weighted by a pre-specified weighting sequence.
The choice of weighting sequence
determines how quickly the algorithm ``forgets" the earlier parameter estimates.
In OEM at time $t$,
\begin{equation} \label{Sgeneral}
\hat{S}^{OEM}_{t}=\gamma_{t} \cdot \tilde{s}_{t} + (1-\gamma_{t}) \cdot \hat{S}^{OEM}_{t-1},
\end{equation}
where $\{\gamma_{k}\}_{k=1,2,...}$ is the chosen weighting sequence, typically of the form
$\gamma_{t}=t^{-c}$ for a chosen $c\in(0.5,1]$ \citep{cappe2009online}. Note that when using lag
$\Delta$, $\gamma_{t}=(t-\Delta)^{-c}$ for $t \geq \Delta$. This update rule ensures that at time $t$,
$\hat{S}^{OEM}$ is a weighted sum of $\{\tilde{s}_{k}\}_{k=\Delta+1,\ldots,t}$ where the term ${\tilde{s}}_k$ has weight
\begin{equation}
\eta_k^t := \gamma_{k} (1-\gamma_{k+1}) \cdots (1-\gamma_{t-1})(1-\gamma_{t}).
\label{weights}
\end{equation}

\noindent \null \hrulefill \\
\textbf{Algorithm 2} Online Expectation Maximization for a simplified autoregressive model\\
\null \hrulefill \\
For time $t \geq 1$:
\begin{enumerate}
\item Simulate and calculate weights of new particles as outlined in Algorithm 1
\item Collect sufficient statistic $\tilde{s}_{t} = \sum_{i=1}^{N} w_t(X^{(i)}_{1:t}) \cdot
                                                    (Y_{t-\Delta}-X^{(i)}_{1:t}(t-\Delta))^2$
\item Update running average of sufficient statistics $\hat{S}^{OEM}_t = \gamma_t \tilde{s}_t +
                                                                         (1-\gamma_t)\hat{S}^{OEM}_{t-1}$
\item Maximize expected likelihood by setting $\hat{\theta}_t := \hat{S}^{OEM}_t$
\end{enumerate}
\hrulefill

Although this method can outperform BEM, its performance remains strong\-ly dependent on the parameter $c$
determining the weighting sequence, and a suboptimal choice can reduce performance by orders of magnitude.
At one extreme, the estimates will depend strongly only on the most recent data, resulting in noisy
parameter estimates and low precision.  At the other extreme, the estimates will average out
stochastic effects but be severely affected by false initial estimates, resulting in
more precise but less accurate estimates. Again, the best choice depends on the model.

A pragmatic approach to the problem of choosing a tuning parameter in OEM
takes inspiration from \cite{polyak1990new}.
In this method, a weight sequence that emphasizes incoming data is used 
to ensure quick initial convergence, while imprecise estimates are avoided
at later iterations by averaging all OEM estimates beyond a threshold $t_0$.
$$
\hat{\theta}^{AVG}_t =
\begin{cases}
\hat{\theta}^{OEM}_t & \text{ for } t < t_0 \\
\frac{1}{t-t_0+1} \sum_{k=t_0}^t \hat{\theta}^{OEM}_k & \text{ for } t \geq t_0.
\end{cases}
$$
Choosing an appropriate threshold $t_0$ can be more straightforward than choosing $c$ for $\gamma_t = t^{-c}$,
but it still requires the user to have an intuition for how the estimates for each parameter will behave.
We will refer to this method as AVG, use $c=0.6$, and set $t_0=50,000$ which is half the total iterations
for our examples.

\subsection{Introspective Online Expectation Maximization}
\label{sec:ioem}

We now introduce IOEM to address the issue of having to pre-specify a weighting sequence $\{\gamma_k\}_{k=1,...}$.
The algorithm is similar to OEM, but instead of pre-specifying $\gamma_{t}$,
we estimate the precision and accuracy in the sufficient statistic updates $\{\tilde{s}_{k}\}_{k=\Delta+1,...,t}$
and use these to determine the next weight $\gamma_{t+1}$.  More precisely, we keep online estimates
of a weighted regression on the dependent variables $\{\tilde{s}_{k}\}_{k=\Delta+1,...,t}$ where
the index $k$ serves as the explanatory variable and the data point $(k,\tilde{s}_{k})$ has weight
\eqref{weights} as before. This weighted regression results in intercept and slope estimates
$\hat{\beta}_0$, $\hat{\beta}_1$, and estimates of their variance $\hat{\sigma}_0^2$, $\hat{\sigma}_1^2$.
We next use these estimates to define a proposed weight as follows:
\begin{equation*}
\gamma_{t+1}^{reg}=\frac{ |\hat{\beta}_1| + \hat{\sigma}_1 }{ \hat{\sigma}_0 },
\end{equation*}
This definition of $\gamma^{reg}_{t+1}$ ensures that a substantial
slope estimate $\hat{\beta}_1$ indicating low accuracy in our previous parameter
estimates will put a large weight on the incoming statistic, improving accuracy.
A large $\hat{\sigma}_0$ reflecting low precision in the estimates will result in a small weight,
so that successive estimates are smoothed out, improving precision.

We do not use standard weighted regression, where the weights are assumed to be inversely
proportional to the variance of the observation, as this assumption is not justified here;
the standard prodecure would lead to biased estimates
of $\hat{\sigma}^2_{0,1}$ and would impact the performance of IOEM.
Instead we assume that observations share an unknown variance, and we use the weights to modulate
the influence of each observation to the estimates of both $\hat{\beta}_{0,1}$ and $\hat{\sigma}^2_{0,1}$.
See Sect.~\ref{sec:weighted_regression} for details.

We impose restrictions on $\gamma_t$ which keep it between the most extreme choices for OEM.
Taken together, the update step for $\gamma$ becomes
\begin{equation}
\gamma_{t+1} = \min\left( (t+1)^{-c}, \max\left( \gamma_{t+1}^{reg}, (t+1)^{-1} \right) \right)
\label{eq:capping}
\end{equation}
where $c>0.5$ is chosen to be very close to $0.5$ and guarantees convergence.
These restrictions ensure that our algorithm satisfies the assumptions of Theorem 1 of \cite{cappe2009line},
namely that $0 < \gamma_t < 1$, $\sum_{t=1}^\infty \gamma_t = \infty$, and $\sum_{t=1}^{\infty} \gamma_t^2 < \infty$.
Hence for any model for which $f$ and $g$ satisfy the assumptions guaranteeing convergence of
the standard OEM estimator, the IOEM algorithm is also guaranteed to converge.
The precise conditions are detailed in Assumption 1, Assumption 2, and Theorem 1 of \cite{cappe2009line}.

\vspace{4mm}
\noindent \null \hrulefill \\
\textbf{Algorithm 3} Introspective Online Expectation Maximization for a simplified autoregressive model\\
\noindent \null \hrulefill \\
For time $t \geq 1$:
\begin{enumerate}
\item Simulate and calculate weights of new particles using SMC with parameter
$ \hat{\theta}_{t-1} $
\item Collect sufficient statistic \\ $\tilde{s}_{t} = \sum_{i=1}^{N} w_t(X^{(i)}_{1:t}) \cdot
                                                    (Y_{t-\Delta}-X^{(i)}_{1:t}(t-\Delta))^2$
\item Maximize expected likelihood by setting \\ $\hat{\theta}_t = \hat{S}^{IOEM}_t := \gamma_t \cdot \tilde{s}_t + (1-\gamma_t) \cdot \hat{S}^{IOEM}_{t-1}$
\item Perform weighted regression on $\tilde{s}$ to calculate $\gamma_{t+1}$
\end{enumerate}
\noindent \null \hrulefill
\vspace{4mm}

\begin{figure}
\centering
\includegraphics[width=\figsize\linewidth]{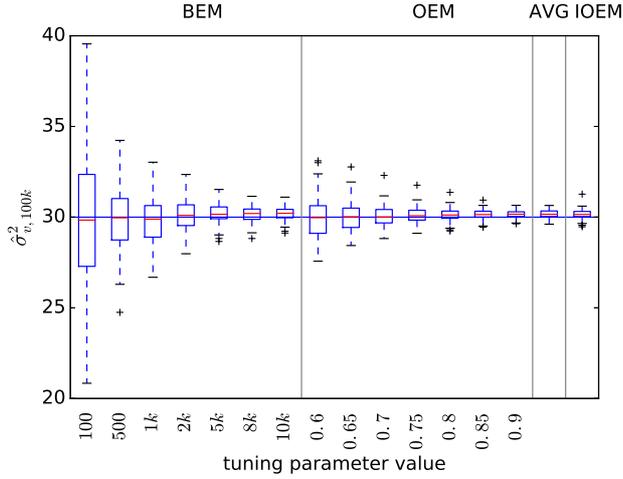}
\captionof{figure}{Comparison of EM methods on simplified AR model with known true parameters $a=.95$,
$\sigma_w=1$, and unknown true $\sigma_v^2=30$, and initial parameter estimate $\sigma_{v,0}^2=20$.
$\hat{\sigma}_{v,100k}^2$ is plotted for 100 replicates, $N=100$}
\label{fig:sar}
\end{figure}

The results of using BEM, OEM, and IOEM to perform parameter inference on model \eqref{ar} with a
wide range of tuning parameters $b$ from $100$ to $10,000$, and $c$ from $0.6$ to $0.9$,
are presented in Figure \ref{fig:sar}. The choice of tuning parameter in BEM and OEM makes a
significant difference to the precision of the estimate even after 100,000 observations.
IOEM was able to recognize that behavior similar to BEM with $b=10,000$ or OEM with $c=0.9$ was optimal.
The accuracy and precision of IOEM are comparable with those of the post-OEM averaging technique (AVG)
with parameters $c=0.6$ and $t_0=50,000$.

The adapting weight sequence $\{\gamma_k\}_{k=1,...}$ sets IOEM apart from OEM.
This formulation of IOEM only works in the setting where $\theta$ has a linear relationship with a
single sufficient statistic (here $\hat{\sigma}_{v,t}^2=\hat{S}_t$) and is meant as an introduction
to some of the ideas involved in IOEM.
The method outlined in Algorithm 3 will not suffice when the function $\Lambda$ mapping the
sufficient statistics to $\theta$ does not have this simple form.
We introduce the general IOEM algorithm in Sect.~\ref{sec:ar} below.

\section{EM Simulations in the Full Autoregressive Model}
\label{sec:ar}

The model of Sect.~\ref{sec:sar} is special in that the sufficient statistic and the parameter of
interest coincide. Generally this is not true, leading to a more involved setup that we explore here.
To this end, we now consider the full noisily-observed autoregressive model AR(1) with master
equations as in \eqref{ar}, but now with unknown parameters $a$, $\sigma_w$, and $\sigma_v$.
We define four sufficient statistics,
\begin{align*}
S_{1,t}&=\mathbb{E}_{X_{1:t} | Y_{1:t}, \theta }\left[\frac{1}{t-1}\sum_{k=1}^{t-1}X_k^2\right],\\
S_{2,t}&=\mathbb{E}_{X_{1:t} | Y_{1:t}, \theta }\left[\frac{1}{t-1}\sum_{k=1}^{t-1}X_k \cdot X_{k+1}\right],\\
S_{3,t}&=\mathbb{E}_{X_{1:t} | Y_{1:t}, \theta }\left[\frac{1}{t-1}\sum_{k=2}^{t}X_k^2\right],\\
S_{4,t}&=\mathbb{E}_{X_{1:t} | Y_{1:t}, \theta }\left[\frac{1}{t}\sum_{k=1}^{t}(Y_k-X_k)^2\right].
\end{align*}
Then, in BEM and OEM, we update the parameter estimates to
\begin{align}
\hat{a}_{t}&=\hat{S}_{2,t}/ \hat{S}_{1,t} \label{a_hat}, \\
\hat{\sigma}_{w,t}&=(\hat{S}_{3,t}-(\hat{S}_{2,t})^2/ \hat{S}_{1,t})^{1/2} \label{sigw_hat},  \\
\hat{\sigma}_{v,t}&=(\hat{S}_{4,t})^{1/2} \label{sigv_hat},
\end{align}
where $\hat{S}_t$ is an approximation of $S_t$.

In most cases, as above, the function $\Lambda$ mapping $\hat{S}_{t}$ to $\hat{\theta}_{t}$
is nonlinear, and requires multiple sufficient statistics as input.
To avoid bias, we want all sufficient statistics that inform one parameter estimate to share a
weight sequence $\{\gamma_{k}\}_{k=1,2,...}$.
We therefore estimate an adapting weight sequence for each parameter independently, by performing
the regression on the level of the parameter estimates (Algorithm 4),
rather than on the level of the sufficient statistics.
We will calculate $\hat{S}_t$ as in OEM \eqref{Sgeneral} using our adapting weight sequence
instead of a user specified weighting sequence.
Because the adapting weight sequence is specific to each parameter,
we will have multiple estimates of certain summary sufficient statistics.
In this case $S_{1,t}$ and $S_{2,t}$ are estimated by $\hat{S}_{1,t}^{a}$ and $\hat{S}_{2,t}^{a}$ for \eqref{a_hat}
and by $\hat{S}_{1,t}^{\sigma_w}$ and $\hat{S}_{2,t}^{\sigma_w}$ for \eqref{sigw_hat}.

Simply regressing on $\hat{\theta}_{1:t}$ with respect to $t$ would correspond to regression on $\hat{S}_{1:t}$,
not $\tilde{s}_{1:t}$. As $\hat{S}$ is a running average, there is a strong correlation between $\hat{S}_{t-1}$
and $\hat{S}_t$ and hence also a strong dependence between $\hat{\theta}_{t-1}$ and $\hat{\theta}_t$.
In order to perform the regression on the parameters we must ``unsmooth" $\hat{\theta}_{1:t}$ to create
pseudo-independent parameter updates $\tilde{\theta}_t$ (see Algorithm 4).
This is accomplished by taking linear combinations,
\begin{equation*}
\tilde{\theta}_{t}:=\frac{1}{\gamma_{t}} \cdot \hat{\theta}_{t} + \left(1-\frac{1}{\gamma_{t}}\right) \cdot \hat{\theta}_{t-1},
\end{equation*}
where the coefficients are chosen so as to minimize the covariance between
successive updates, justifying the term pseudo-independent.
The resulting updates correspond with the unsmoothed sufficient statistics
updates $\tilde{s}_{t}$ used in Sect.~\ref{sec:ioem}. See Sect.~\ref{sec:parameter_updates} for
further details on this step.

\vspace{4mm}
\vbox{
\noindent \null \hrulefill \\
\textbf{Algorithm 4} Introspective Online Expectation Maximization in the general model\\
\noindent \null \hrulefill \\
For time $t \geq 1$:
\begin{enumerate}
\item Simulate and calculate weights of new particles using SMC with parameter
$ \hat{\theta}^{IOEM}_{t-1} $
\item Collect sufficient statistics $\tilde{s}_{t}$
\item Update running average of sufficient statistics \\
$\hat{S}_t = \gamma_t \tilde{s}_t + (1-\gamma_t) \hat{S}_{t-1}$
\item Maximize expected likelihood $\hat{\theta}_t = \Lambda(\hat{S}_t)$
\item Create pseudo-independent parameter updates \\
$\tilde{\theta}_t = \frac{1}{\gamma_t} \cdot \hat{\theta}_t + (1 - \frac{1}{\gamma_t}) \cdot \hat{\theta}_{t-1}$
\item Perform weighted regression on $\tilde{\theta}$ to calculate $\gamma_{t+1}$
\end{enumerate}
\noindent \null \hrulefill
}

\begin{figure}
\centering
\includegraphics[width=\figsize\linewidth]{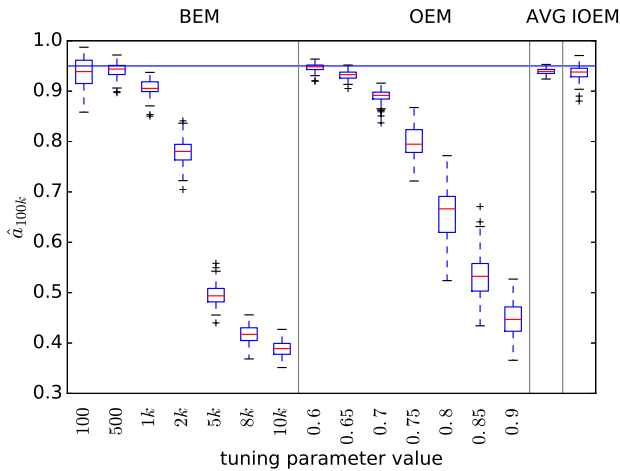}
\captionof{figure}{Comparison of EM methods on full autoregressive model with unknown true parameters
$a=0.95$, $\sigma_{w}=1$, $\sigma_{v}=5.5$  and inital parameters
$a_{0}=0.8$, $\sigma_{w,0}=3$, $\sigma_{v,0}=1$. $\hat{a}_t$ at $t=100,000$
is plotted for 100 replicates, $N=100$}
\label{fig:ar}       
\end{figure}

Estimates for the $a$ parameter under different EM methods are presented in Fig.\ \ref{fig:ar}; for
the other parameter inferences see Sect.~\ref{sec:figures}, Fig. \ref{fig:sup1}.
In the AR(1) model, IOEM outperforms most other EM methods when estimating the $a$ parameter.
It is worth noting that in this case, OEM with $c=0.6$ substantially outperforms OEM with $c=0.9$.
This is a result of the bad initial estimates. OEM with $c=0.6$ forgets the earlier
simulations much faster than OEM with $c=0.9$ and hence is able to move its estimates of $a$,
$\sigma_w$, and $\sigma_v$ much more quickly.
Here IOEM recognizes that it should have similar behavior to OEM with $c=0.6$,
whereas in the inference displayed in Figure \ref{fig:sar} IOEM chose behavior similar to OEM with $c=0.9$.
IOEM can indeed adapt to the model.

\section{EM Simulations in a Two-Dimensional AR Model}
\label{sec:2dar}

Now we investigate a model with a larger number of parameters and varying accuracy
of initial parameter estimates. IOEM's main advantage over OEM is its ability to adapt to each parameter independently.
To highlight this, we applied IOEM to a simple 2-dimensional autoregressive model.
For this model we consider the sequences $\{Y^A,Y^B\}_{1:t}$ as observed, while
$\{X^A,X^B\}_{1:t}$ are unobserved, where
\begin{align}
X^{A}_{t}&=a^{A} X^{A}_{t-1}+ \sigma^{A}_{w} W^{A}_{t},
& X^{B}_{t}&=a^{B} X^{B}_{t-1}+ \sigma^{B}_{w} W^{B}_{t}, \nonumber \\
Y^{A}_{t}&=X^{A}_{t} + \sigma_{v} V^{A}_{t},
& Y^{B}_{t}&=X^{B}_{t} + \sigma_{v} V^{B}_{t}.
\label{twoar}
\end{align}
Note that $Y^A$ and $Y^B$ are uncoupled, and that their master equation have
independent parameters except for a shared parameter $\sigma_v$.
By giving component $A$ good initial estimates and $B$ bad initial estimates,
we can see how the different EM methods cope with a combination of accurate and
inaccurate initializations. IOEM is able to identify the set with good initial estimates
($a^A,\sigma^{A}_{w}$) and quickly start smoothing out noise.
To IOEM, the other parameters appear to not have converged ($\sigma^{B}_{w}$
and $\sigma_{v}$ because they are at the wrong value,
$a^B$ because it will be changing to compensate for $\sigma^{B}_{w}$ and $\sigma_{v}$).

OEM with $c=0.6$ and OEM with $c=0.9$ both suffer in this model as they are
both well suited to parameter estimation in one of the components, but not the other.
IOEM on the other hand is able to capture the best of both worlds, striving for
precision in component A and initially foregoing precision in favour of accuracy in component B.

\begin{figure}
\centering
\includegraphics[width=\figsize\linewidth]{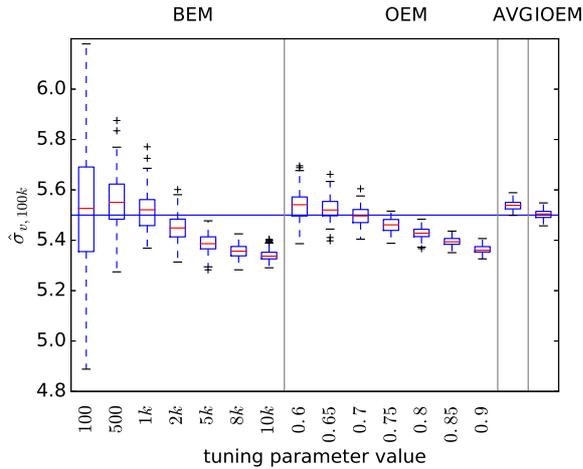}
\captionof{figure}{Comparison of EM methods on 2-dimensional autoregressive model with true
parameters $a^A=0.95$, $\sigma^{A}_{w}=1$, $\sigma_{v}=5.5$, $a^B=0.95$, $\sigma^{B}_{w}=1$
and inital parameters $a^{A}_{0}=0.95$, $\sigma^{A}_{w,0}=1$, $\sigma_{v,0}=3$, $a^{B}_{0}=0.95$,
$\sigma^{B}_{w,0}=3$. $\hat{\sigma}_{v,t}$ at $t=100,000$ is plotted for 100 replicates, $N=100$}
\label{fig:2dar}       
\end{figure}

Figure \ref{fig:2dar} shows the inference of $\sigma_v$, which due to its dependence on components A and B,
suffers the most from a blanket choice of tuning parameter in BEM or OEM. The inference of the other
parameters and comparisons with a different choice of AVG threshold are shown in Sect.~\ref{sec:figures},
figures \ref{fig:sup3}-\ref{fig:sup6}.

\section{Stochastic volatility model}
\label{sec:sv}

\begin{figure}
\centering
\includegraphics[width=\figsize\linewidth]{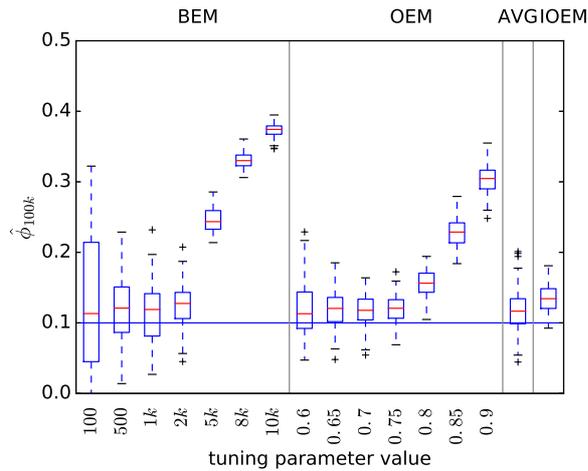}
\captionof{figure}{Estimates of  in stochastic volatility model}
\label{fig:sv}
\end{figure}

The previous sections have demonstrated IOEM is comparable to choosing the optimal tuning parameter
in OEM or BEM in certain models. However, the models shown have all been based on the noisily
observed autoregressive model, which is a linear Gaussian case where in practice analytic techniques
would be prefered over SAEM. We now examine the behaviour of these algorithms when inferring the
parameters of a non-linear stochastic volatility model defined by transition and emission densities

\begin{align*}
f(x_t|x_{t-1}) &= (2 \pi \sigma^2)^{-1/2} \exp\Big\{-\frac{(x_t - \phi x_{t-1})^2}{2 \sigma^2}\Big\}, \\
g(y_t|x_t) &= (2 \pi \beta^2 e^{x_t} )^{-1/2} \exp\Big\{-\frac{ 1 }{ 2 \beta^2 e^{x_t} } y^2_t \Big\}.
\end{align*}
We define four summary sufficient statistics,
\begin{align*}
S_{1,t}&=\mathbb{E}_{X_{1:t} | Y_{1:t}, \theta }\left[\frac{1}{t-1}\sum_{k=1}^{t-1}X_k \cdot X_{k+1}\right],\\
S_{2,t}&=\mathbb{E}_{X_{1:t} | Y_{1:t}, \theta }\left[\frac{1}{t-1}\sum_{k=1}^{t-1}X_k^2\right],\\
S_{3,t}&=\mathbb{E}_{X_{1:t} | Y_{1:t}, \theta }\left[\frac{1}{t-1}\sum_{k=2}^{t}X_k^2\right],\\
S_{4,t}&=\mathbb{E}_{X_{1:t} | Y_{1:t}, \theta }\left[\frac{1}{t}\sum_{k=1}^{t} e^{-X_k} \cdot Y_k^2 \right].
\end{align*}
Then the set of parameters that maximises the likelihood at step $t$ are
\begin{align}
\hat{\phi}_{t}&=\hat{S}_{1,t}/ \hat{S}_{2,t} \label{phi_hat}, \\
\hat{\sigma}_{t}&=(\hat{S}_{3,t}-(\hat{S}_{1,t})^2/ \hat{S}_{2,t})^{1/2} \label{sig_hat},  \\
\hat{\beta}_{t}&=(\hat{S}_{4,t})^{1/2} \label{beta_hat},
\end{align}
Again IOEM results in similar estimates to the optimal BEM/OEM and the online averaging technique 
with a well-chosen threshold (see Fig.\ \ref{fig:sv} and Sect.~\ref{sec:figures}, Fig.\ \ref{fig:sup2}).

\section{Conclusion}
\label{sec:discussion}

Stochastic Approximation EM is a general and effective technique for estimating parameters in the context of SMC.
However, convergence can be slow, and improving convergence speed is of particular interest in this setting.
We have shown that IOEM produces accurate and precise parameter estimates when applied to continuous
state-space models. Across models, and across varying levels of accuracy of the intial estimates, the efficiency
of IOEM matches that of BEM/OEM with the optimal choice of tuning parameter.
The AVG procedure also shows good behaviour, but like BEM/OEM it has tuning parameters, and when
these are chosen suboptimally performance is not as good as IOEM (Figs.~\ref{fig:sup5}-\ref{fig:sup6}).
In addition, BEM/OEM/AVG all make use of a single learning schedule $\{\gamma_{(k)}\}$, 
and for more complex models a single learning schedule generally cannot achieve optimal convergence rates for all 
parameters, as we have shown for the 2-dimensional AR example.

IOEM finds parameter-specific learning schedules, resulting in better performance than standard methods 
with a single learning rate parameter are able to achieve.
IOEM can be applied with minimal prior knowledge of the model's behavior, and requires no user supervision,
while retaining the convergence guarantees of BEM/OEM, therefore providing an efficient, practical 
approach to parameter estimation in SMC methods.

\Urlmuskip=0mu plus 1mu\relax
\bibliography{ms}
\bibliographystyle{apalike}

\bigskip
\begin{center}
{\large\bf SUPPLEMENTAL MATERIALS}
\end{center}
\section{Supplemental text}
\label{sec:supplement}
\begin{subsection}{Fixed-lag technique}
\label{sec:lag}

Our fixed-lag technique is slightly different than that proposed in the literature
\citep{cappe2005use,olsson2008sequential}.
Compared to the existing approach it uses less intermediate storage.
Recall that the approximation we aim to evaluate is
\begin{equation*}
  \hat{S}_t = \sum_i
  w_t(X_{1:t}^{(i)}) \cdot
  \sum_{u=1}^t s_u( X_{1:t}^{(i)}(u), Y(u) ),
\end{equation*}
where the sufficient statistic is written explicitly as a sum over the path traced out by
the particle $X_{1:t}^{(i)}$.  The drawback is that for $u \ll t$ the paths will have
collapsed due to resampling, increasing the variance for those contributions to $S$.
The solution proposed in \citet{cappe2005use} is to use instead the approximation
\begin{align*}
  \hat{S}_t \approx \sum_i
  \bigg(&
  \sum_{u=1}^{t-\Delta} w_{u+\Delta}(X_{1:u+\Delta}^{(i)}) s_u( X_{1:u+\Delta}^{(i)}(u), Y(u) ) \\
  &+ w_t(X_{1:t}^{(i)}) \sum_{u=t-\Delta+1}^{t} s_u( X_{1:t}^{(i)}(u), Y(u) ) \bigg).
\end{align*}
This requires storing the quantities
\begin{equation*}
  \{ s_u(X_{1:u+\Delta}^{(i)}(u)), Y(u) \}_{u=t-\Delta,...,t}
\end{equation*}
for each
sufficient statistic and each particle.
This storage can be expensive if large numbers of sufficient statistics are tracked.
Instead, at iteration $t$ we use the approximation
\begin{equation*}
  \hat{S}_{t} \approx \sum_{u=1}^{t-\Delta} \sum_i w_{u+\Delta}(X^{(i)}_{1:u+\Delta}) s_u(X^{(i)}_{1:u+\Delta}(u),Y(u)).
\end{equation*}
By disregarding terms involving $s_u$ for $u>t-\Delta$ and switching the summation in this way,
we can now update $\hat{S}$ at each iteration
by adding the contribution of the current particles to a single summary statistic at a distance $\Delta$,
without requiring per-particle storage other than each particle's recent history.
\end{subsection}

\begin{subsection}{Weighted regression}
\label{sec:weighted_regression}

The term ``weighted regression" usually refers to regression where the errors are independent and
normally distributed with zero mean and
known variance (up to a multiplicative constant), and the data is weighted inversely proportionally to its variance.
In our case, the data is assumed to drift, contributing an additional, non-independent term to the error.
Weights are used to only focus on recent data where the drift contributes an error
of the same order of magnitude as the normally distributed noise, while discounting the impact of data points
further away.
In this setup we are interested both in estimating the regression coefficients, and the error in these estimates.

Perry Kaufman's adaptive moving average (AMA) \citep{kaufman1995smarter} is a similar averaging
technique which reacts to the trends and volatility (jointly referred to as the behavior) of the sequence.
The difference lies in the measure of the behavior. AMA relies on a user specified window length $n$.
The $n$ most recent data points are used to measure the behavior. This would be equivalent to
using equally-weighted linear regression over the last $n$ points. By using weighted regression,
the contribution of points to the behavior measures is also influenced by the previously observed behavior.
For example, a sharp trend will effectively employ a smaller $n$ value as we have lost interest
in the behavior before that trend.

Let $X$ be the $2\times n$ matrix consisting of a column of $1$s and a column with the dependent variable, let $y$ be
the vector of observations, let $\beta$ be the two coefficients, and $\epsilon$ the vector of errors, with
$\epsilon_k \sim N(0,\sigma^2)$.  Finally let $w$ be a vector of weights.  We estimate $\beta$ by minimizing
\begin{align*}
  s^2 
   &= (X_w\beta - y_w)^\top (X_w\beta - y_w),
\end{align*}
where $X_w$ and $y_w$ are defined as
\begin{equation}
     X_w:=\begin{bmatrix}
         w_1 & w_1 \cdot (-n+1) \\
         \vdots & \vdots \\
         w_n & w_n \cdot 0
        \end{bmatrix};\qquad
     y_w:=\begin{bmatrix}
         w_1 \cdot y_1 \\
         \vdots \\
         w_n \cdot y_n
        \end{bmatrix}. \nonumber
\end{equation}
Setting the derivative $\partial s^2/\partial \beta = 2(X_w\beta - y_w)^\top X_w$
to zero and solving for $\beta$ results in the standard estimator for weighted regression
\begin{equation*}
  \hat{\beta} = (X^\top_w X_w)^{-1}X^\top_w y_w,
\end{equation*}
or more explicitly
\begin{align}
\hat{\beta}_1 &= \frac{ (\sum w_k^2 x_{2k} y_k) - (\sum w_k^2 x_{2k} ) (\sum w
_k^2 y_k) } { (\sum w_k^2 x_{2k}^2) - (\sum w_k^2 x_{2k})^2 }, \nonumber \\
\hat{\beta}_0 &= \frac{ (\sum w_k^2 x_{2k}^2 ) (\sum w_k^2 y_k) - (\sum w_k^2 x_{2k} y_k )( \sum w_k^2 x_{2k} ) }{ (\sum w_k^2 x_{2k}^2) - (\sum w_k^2 x_{2k})^2 }. \nonumber
\end{align}
From this expression we can see that $\hat{\beta}$ can be updated in 
an online manner as $k$ increases simply by updating the above summations.
The variance in $\hat{\beta}$ can be estimated as follows:
\begin{align*}
  \var \hat{\beta} &= \var  (X^\top_w X_w)^{-1} X^\top_w y_w \\
  &= \var (X^\top_w X_w)^{-1} X^\top_w \epsilon_w \\
  &= E \left[
  (X^\top_w X_w)^{-1} X^\top_w \epsilon_w \epsilon_w^\top X_w (X^\top_w X_w)^{-1}
    \right]
  \\
  &= (X^\top_w X_w)^{-1} X^\top_w \diag( w_k^2 \sigma^2 ) X_w (X^\top_w X_w)^{-1}.
\end{align*}
If $w_k^2=1$ this simplifies to the usual $\var \hat{\beta} = \sigma^2 (X^\top X)^{-1}$.
Writing out the expression for $\var \hat{\beta}$ explicitly shows that
it is again possible to find online updates for the relevant terms.
\end{subsection}

\begin{subsection}{Pseudo-independent parameter updates}
\label{sec:parameter_updates}

In order to perform our regression on the level of the parameters, we need to map from
$\tilde{s}^{(t)}$ to $\hat{S}^{(t)}$ and then to $\hat{\theta}^{(t)}$.
We do not wish to regress on $\hat{\theta}^{(1:t)}$, as $\hat{\theta}^{(t-1)}$ and $\hat{\theta}^{(t)}$
are highly correlated. Instead we want a sequence defined in the parameter space where the
correlations resemble those in $\tilde{s}^{(1:t)}$.
We define this sequence as
\begin{equation*}
\tilde{\theta}_{t}:=\frac{1}{\gamma_{t}}\hat{\theta}_{t}+\left(\frac{\gamma_{t}-1}{\gamma_{t}}\right)\hat{\theta}_{t-1}.
\end{equation*}
Here we show that $\tilde{\theta}_{i}$ and $\tilde{\theta}_{j}$ are uncorrelated for all $i\neq j$,
under the assumption that  $\tilde{s}_{i}$ and $\tilde{s}_{j}$ are uncorrelated ($i\neq j$).
Define $\{\eta_k^t \}_{k=0,...,t}$ to be the sequence that satisfies
$\hat{S}_t = \sum_{k=0}^t \eta_k^t \tilde{s}_k$ and $\sum_{k=0}^t \eta_k^t = 1$.
Note that $\eta_t^t=\gamma_t$, $\eta_{t-1}^t=\gamma_{t-1} (1-\gamma_t)$, and so on.  Now,
\begin{align}
\cov (\tilde{\theta}_{i}, \tilde{\theta}_{j}) 
=\,& \cov \biggl( \frac{1}{ \gamma_{i} } \hat{\theta}_{i} + \frac{ \gamma_{i}-1 }{ \gamma_{i} } \hat{\theta}_{i-1} , \frac{1}{ \gamma_{j} } \hat{\theta}_{j} + \frac{ \gamma_{j}-1 }{ \gamma_{j} } \hat{\theta}_{j-1} \biggr) \nonumber \\
=\,& \frac{1}{\gamma_i \gamma_j} \cov ( \hat{\theta}_i,\hat{\theta}_j ) \nonumber \\
&+ \frac{1}{\gamma_j} \left(1-\frac{1}{\gamma_i}\right) \cov ( \hat{\theta}_{i-1}, \hat{\theta}_j ) \nonumber \\
&+ \frac{1}{\gamma_i} \left(1-\frac{1}{\gamma_j}\right) \cov ( \hat{\theta}_i, \hat{\theta}_{j-1} ) \nonumber \\
&+ \left(1-\frac{1}{\gamma_i}\right) \left(1-\frac{1}{\gamma_j}\right) \cov ( \hat{\theta}_{i-1}, \hat{\theta}_{j-1} ). \label{parts}
\end{align}

Writing $\hat{\theta}_i=f_0+f_1\sum_{k=0}^i \eta_k^i \tilde{s}_k$ and recalling that
\begin{equation*}
\cov(\tilde{s}_i,\tilde{s}_j) =
\begin{cases}
0, & \text{if } i \neq j \\
\sigma_i^2, & \text{if } i=j,
\end{cases}
\end{equation*}
it follows that
\begin{align}
\cov(\hat{\theta}_i,\hat{\theta}_j) &= \cov \Big ( f_1 \sum_{k=0}^i \eta_k^i \tilde{s}_k,f_1 \sum_{k=0}^j \eta_k^j \tilde{s}_k \Big ) \nonumber \\
&= \sum_{k=0}^i f_1^2 \eta_k^i \eta_k^j \sigma_i^2, \nonumber
\end{align}
 for $i<j$.
Substituting into the four terms of (\refeq{parts}) yields
\begin{align*}
  \cov(\tilde{\theta}_i,\tilde{\theta}_j) =\,
 & \frac{1}{\gamma_i \gamma_j} \sum_{k=0}^i f_1^2 \eta_k^i \eta_k^j \sigma_k^2 \\
 &+ \frac{1}{\gamma_j } \left( \frac{\gamma_i-1}{\gamma_i} \right) \sum_{k=0}^{i-1} f_1^2 \eta_k^{i-1} \eta_k^j \sigma_k^2 \\
 &+ \frac{1}{\gamma_i } \left( \frac{\gamma_j-1}{\gamma_j} \right) \sum_{k=0}^{i} f_1^2 \eta_k^{i} \eta_k^{j-1} \sigma_k^2 \\
 &+\left( \frac{\gamma_i-1}{\gamma_i } \right) \left( \frac{\gamma_j-1}{\gamma_j} \right) \sum_{k=0}^{i-1} f_1^2 \eta_k^{i-1} \eta_k^{j-1} \sigma_k^2.
\end{align*}


If we define 
\begin{align}
a := f_1^2 \eta_i^i \eta_i^{j-1} \sigma_i^2, \nonumber
\end{align}
\begin{align}
b := \sum_{k=0}^{i-1} f_1^2 \eta_k^{i-1} \eta_k^{j-1} \sigma_k^2, \nonumber
\end{align}
and note that
\begin{align}
\eta_k^j=(1-\gamma_j) \eta_k^{j-1} \text{ for all } k<j, \nonumber
\end{align}
then
\begin{align}
\cov  (\tilde{\theta}_i,\tilde{\theta}_j) =\,& 
 \frac{1}{\gamma_i \gamma_j} (1-\gamma_j) a + \frac{1}{\gamma_i \gamma_j} (1-\gamma_i) (1-\gamma_j) b \nonumber \\
&+  \frac{1}{\gamma_j } \left( \frac{\gamma_i-1}{\gamma_i} \right) (1-\gamma_j) b  \nonumber \\
&+ \frac{1}{\gamma_i } \left( \frac{\gamma_j-1}{\gamma_j} \right) a + \frac{1}{\gamma_i } \left( \frac{\gamma_j-1}{\gamma_j} \right) (1-\gamma_i) b \nonumber \\
&+ \left( \frac{\gamma_i-1}{\gamma_i } \right) \left( \frac{\gamma_j-1}{\gamma_j} \right) b \nonumber \\
=\,& 0. \nonumber
\end{align}
Hence, if $\tilde{s}_{i}$ and $\tilde{s}_{j}$ are independent for all $i\neq j$,
then $\tilde{\theta}_i$ and $\tilde{\theta}_j$ are uncorrelated ($i \neq j$), justifying the
term ``pseudo-independent updates'' for $\tilde{\theta}_i$.

\end{subsection}


\begin{subsection}{Notation reference}
\label{sec:table}


\begin{table}[h]
\begin{tabular}{|l|l|l|}
\hline
notation & \multicolumn{1}{p{6.01cm}|}{meaning} & \multicolumn{1}{p{2.5cm}|}{associated methods} \\
\hline
$\theta$ & \multicolumn{1}{p{6.01cm}|}{\raggedright true parameter} & all \\
\hline
$\hat{\theta}_t$ & \multicolumn{1}{p{6.01cm}|}{\raggedright parameter estimate at time $t$} & all \\
\hline
$\tilde{\theta}_t$ & \multicolumn{1}{p{6.01cm}|}{\raggedright pseudo-independent parameter update} & IOEM \\
\hline
$\tilde{s}_t$ & \multicolumn{1}{p{6.01cm}|}{\raggedright sufficient statistic update at time $t$} & all \\
\hline
$\hat{S}_t$ & \multicolumn{1}{p{6.01cm}|}{\raggedright summary sufficient statistic from averaging $\tilde{s}$} & all \\
\hline
$N$ & \multicolumn{1}{p{6.01cm}|}{\raggedright number of particles} & all \\
\hline
$\Delta$ & \multicolumn{1}{p{6.01cm}|}{\raggedright lag of fixed-lag technique} & all \\
\hline
$\hat{\beta}_0$ & \multicolumn{1}{p{6.01cm}|}{\raggedright regression intercept ML estimate} & IOEM \\
\hline
$\hat{\beta}_1$ & \multicolumn{1}{p{6.01cm}|}{\raggedright regression slope ML estimate} & IOEM \\
\hline
$\hat{\sigma}_0^2$ & \multicolumn{1}{p{6.01cm}|}{\raggedright variance of regression intercept ML estimate} & IOEM \\
\hline
$\hat{\sigma}_1^2$ & \multicolumn{1}{p{6.01cm}|}{\raggedright variance of regression slope ML estimate} & IOEM \\
\hline
\end{tabular}
\captionof{table}{Notation used in this paper}
\label{table:notation}
\end{table}

\end{subsection}

\begin{subsection}{Supplementary figures}
\label{sec:figures}

\begin{figure*}[!h]
\includegraphics[width=0.328\linewidth]{figures/ar_a_100k_final.eps}
\includegraphics[width=0.328\linewidth]{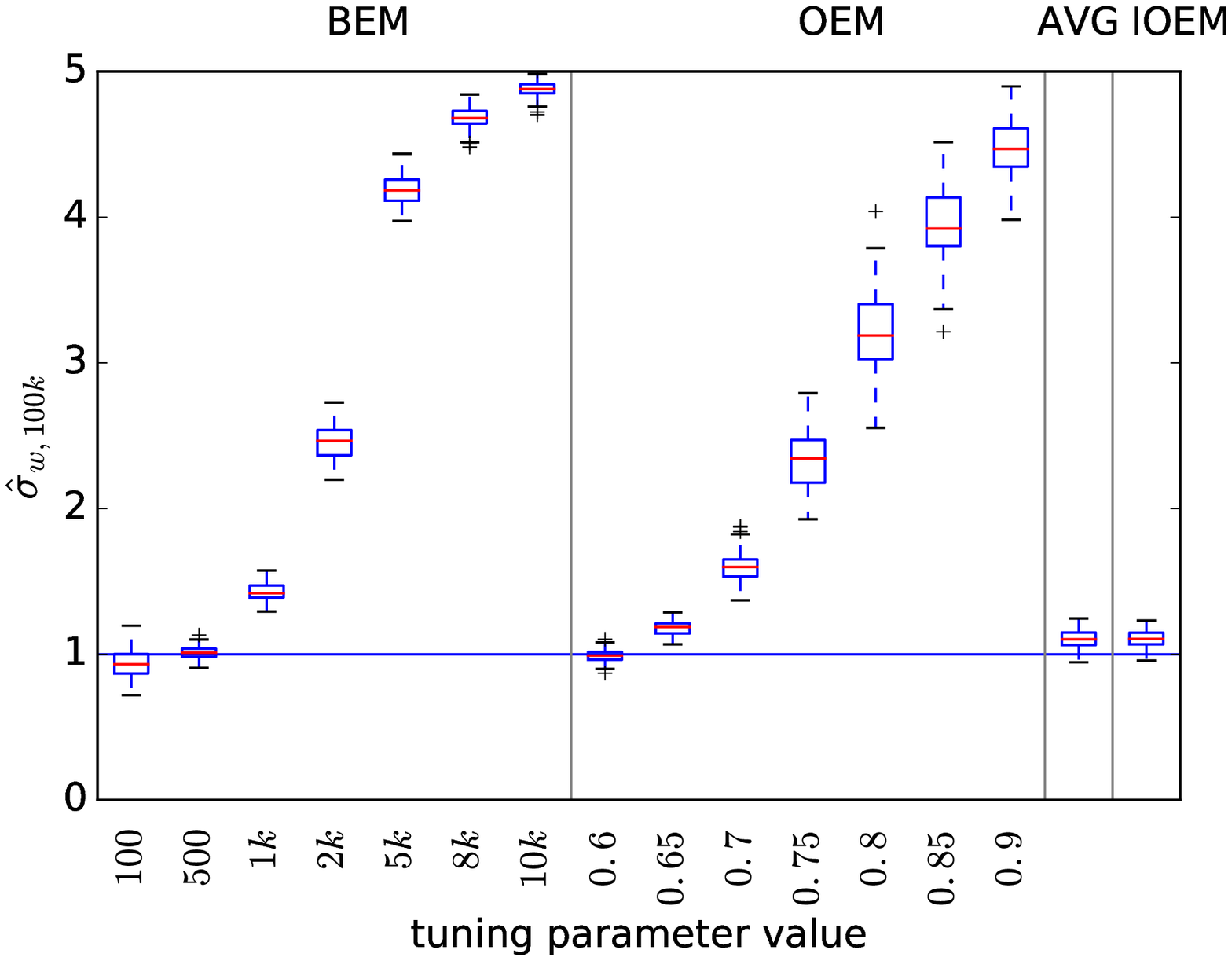}
\includegraphics[width=0.328\linewidth]{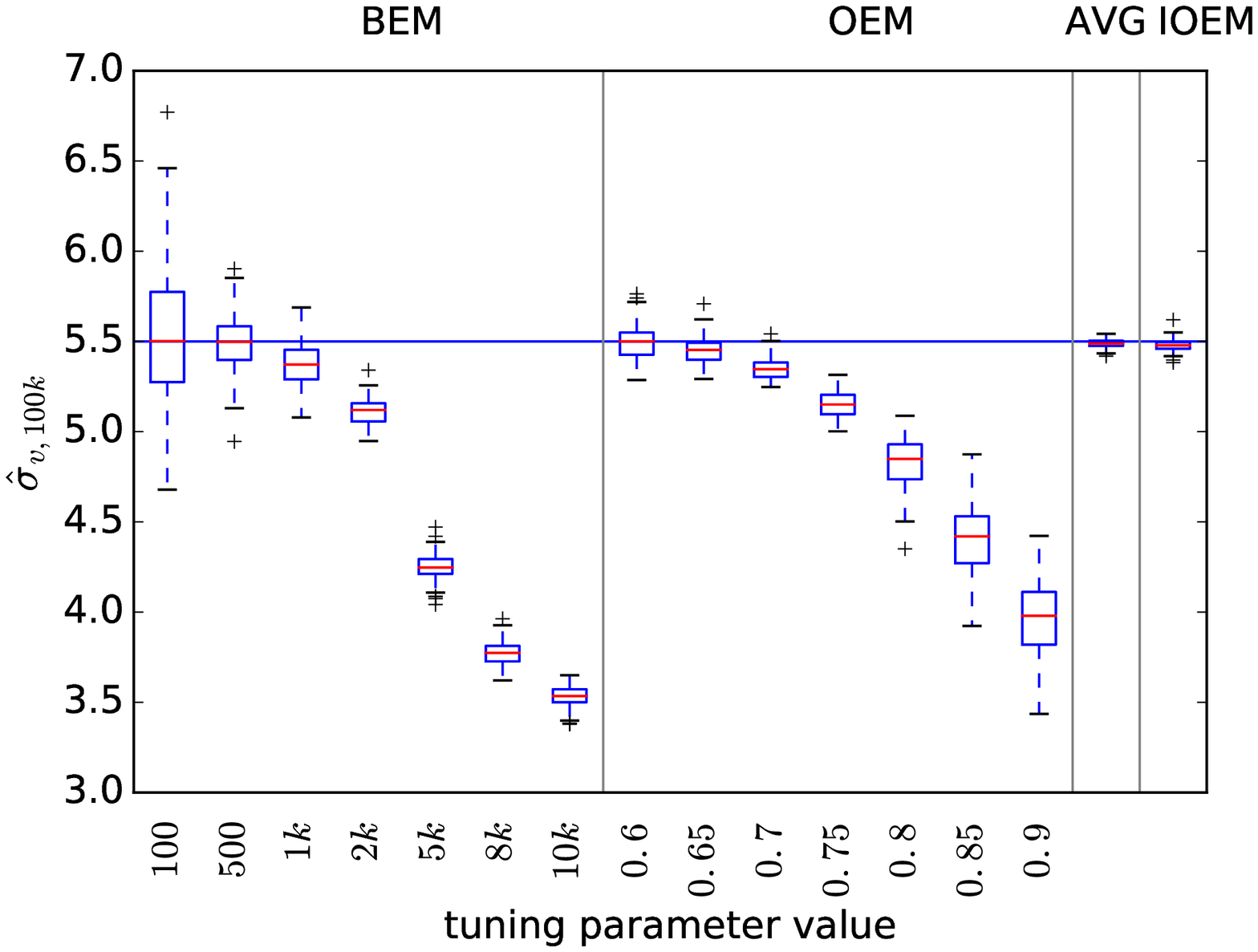}
\caption{Comparison of EM methods on full autoregressive model
with unknown true parameters $a=0.95$, $\sigma_{w}=1$, $\sigma_{v}=5.5$  and inital parameters
$a_{0}=0.8$, $\sigma_{w,0}=3$, $\sigma_{v,0}=1$. Parameter estimates at $t=100,000$
are plotted for 100 replicates, $N=100$}
\label{fig:sup1}
\end{figure*}

\begin{figure*}[!h]
\includegraphics[width=0.328\linewidth]{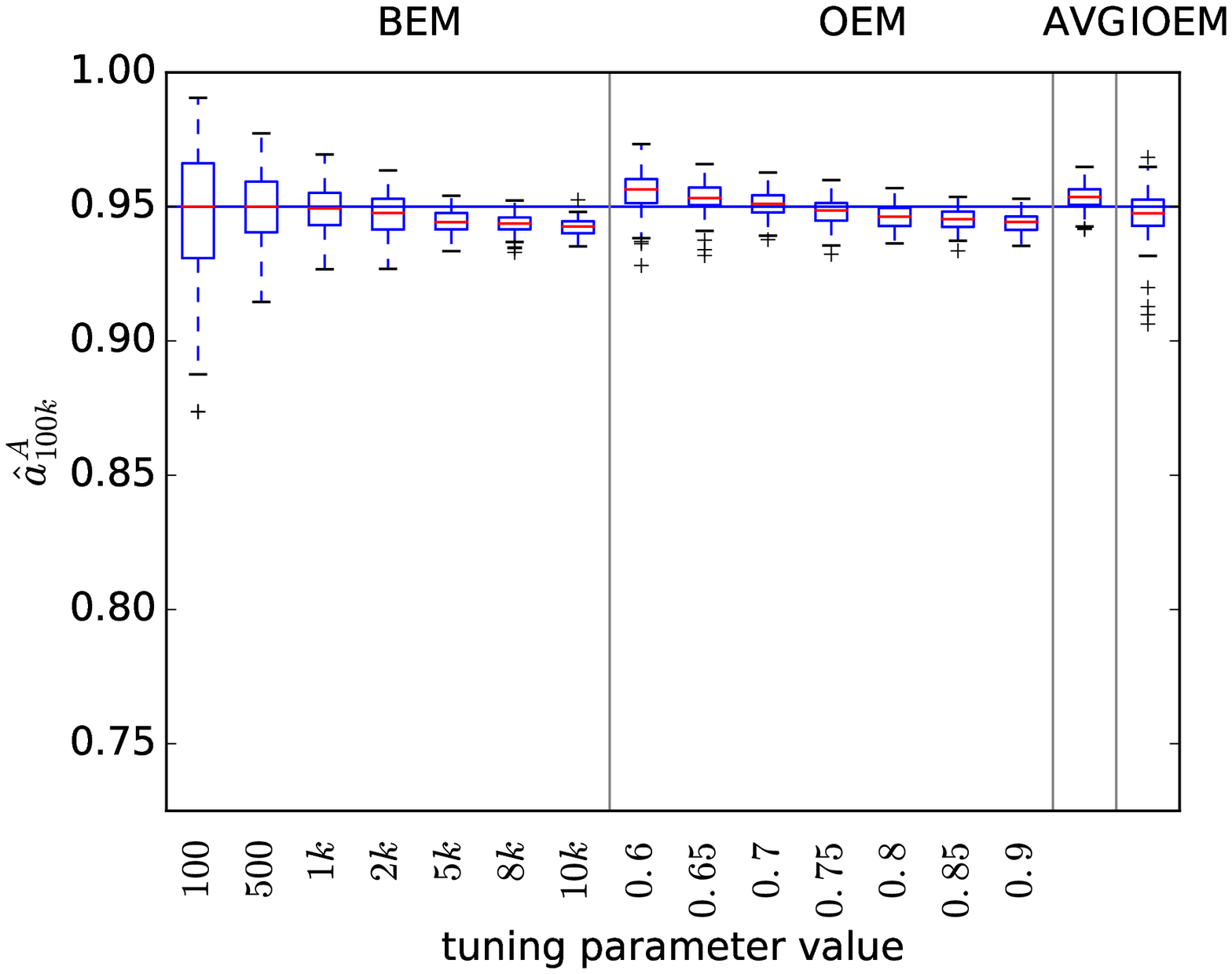}
\includegraphics[width=0.328\linewidth]{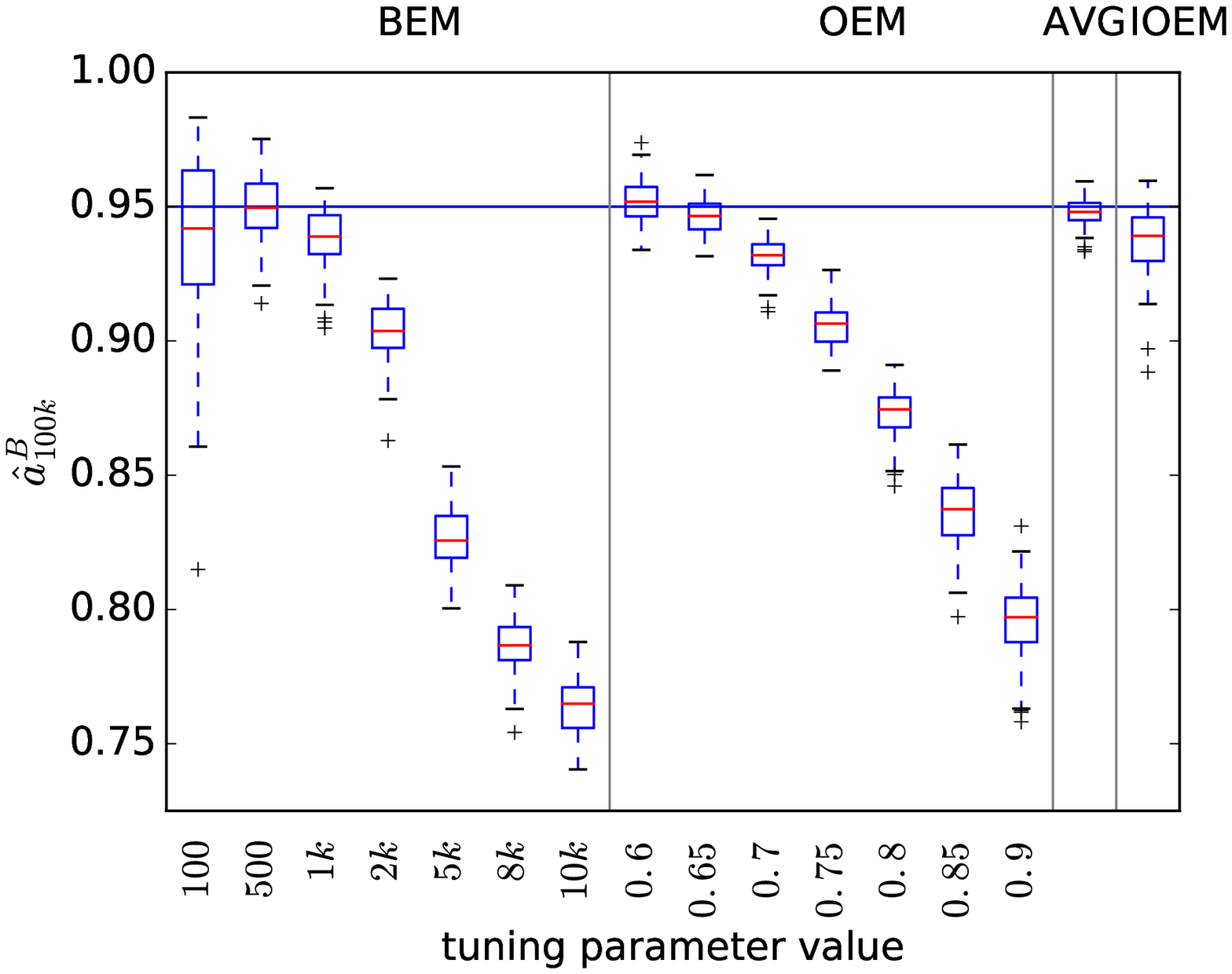}
\\
\includegraphics[width=0.328\linewidth]{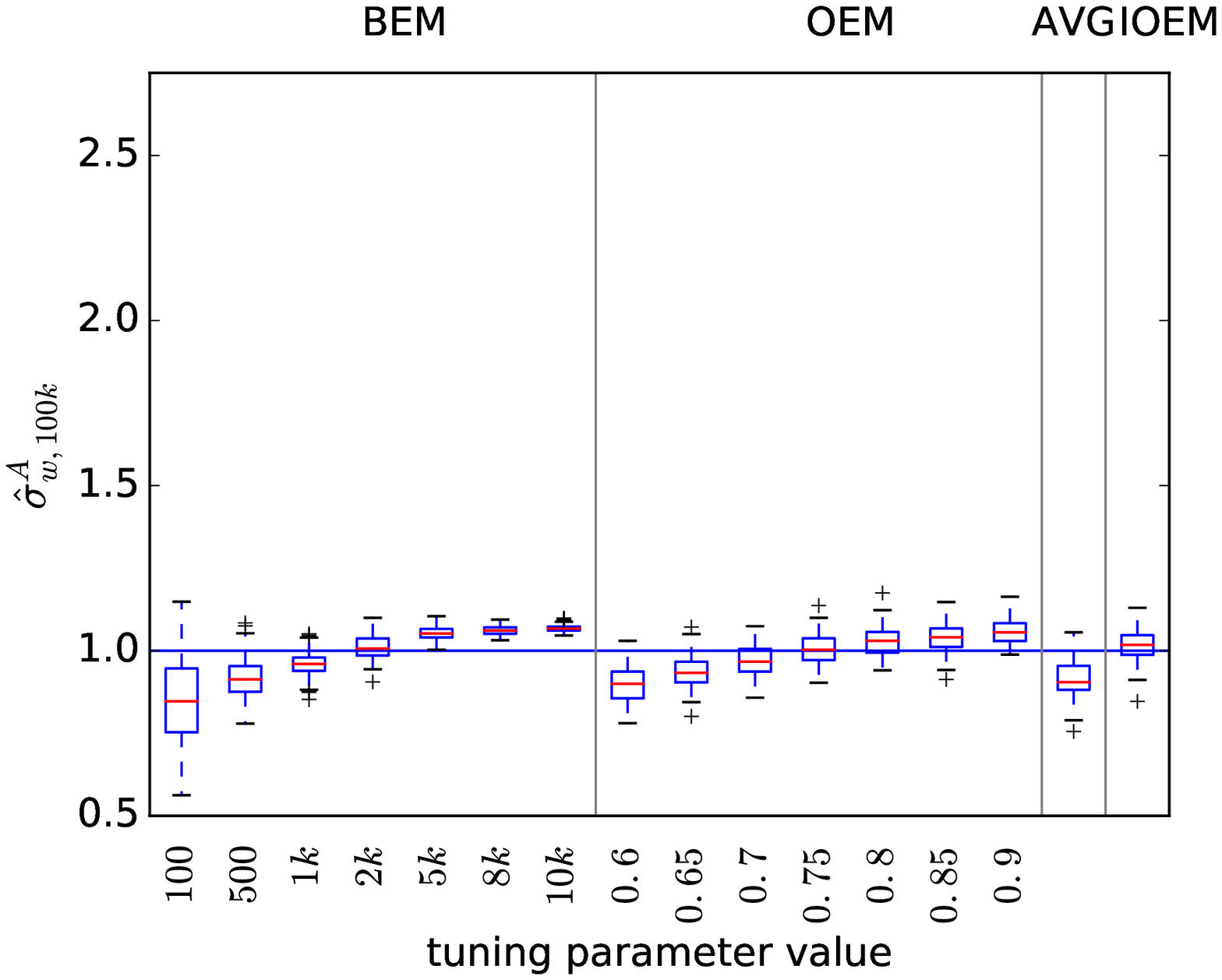}
\includegraphics[width=0.328\linewidth]{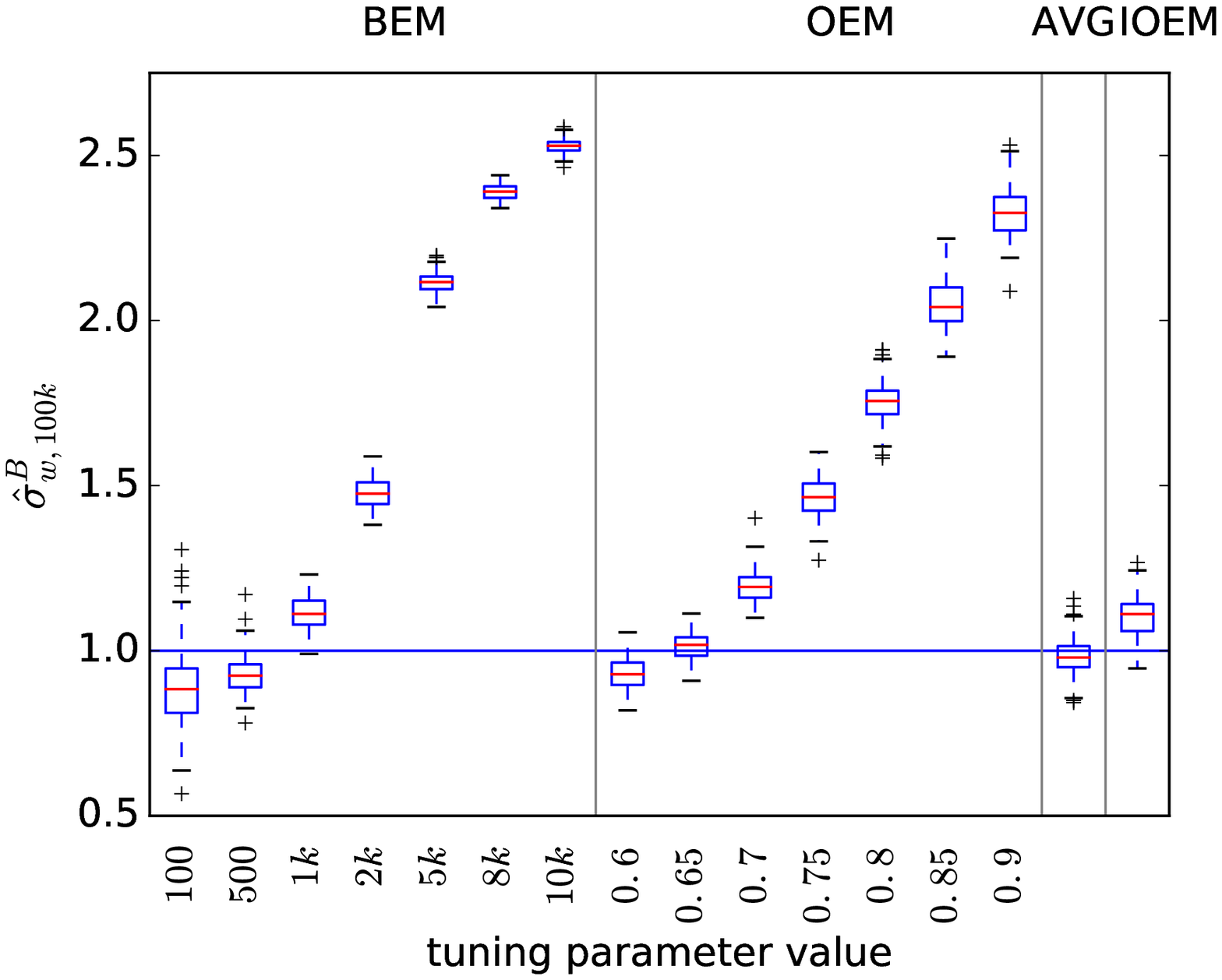}
\includegraphics[width=0.328\linewidth]{figures/twoar_sigv_100k_final.eps}
\caption{Comparison of EM methods on 2-dimensional autoregressive model with true
parameters $a^A=0.95$, $\sigma^{A}_{w}=1$, $\sigma_{v}=5.5$, $a^B=0.95$, $\sigma^{B}_{w}=1$
and inital parameters $a^{A}_{0}=0.95$, $\sigma^{A}_{w,0}=1$, $\sigma_{v,0}=3$, $a^{B}_{0}=0.95$,
$\sigma^{B}_{w,0}=3$. Parameter estimates at $t=100,000$ are plotted for 100 replicates, $N=100$}
\label{fig:sup3}
\end{figure*}

\begin{figure*}[!h]
\includegraphics[width=.99\linewidth]{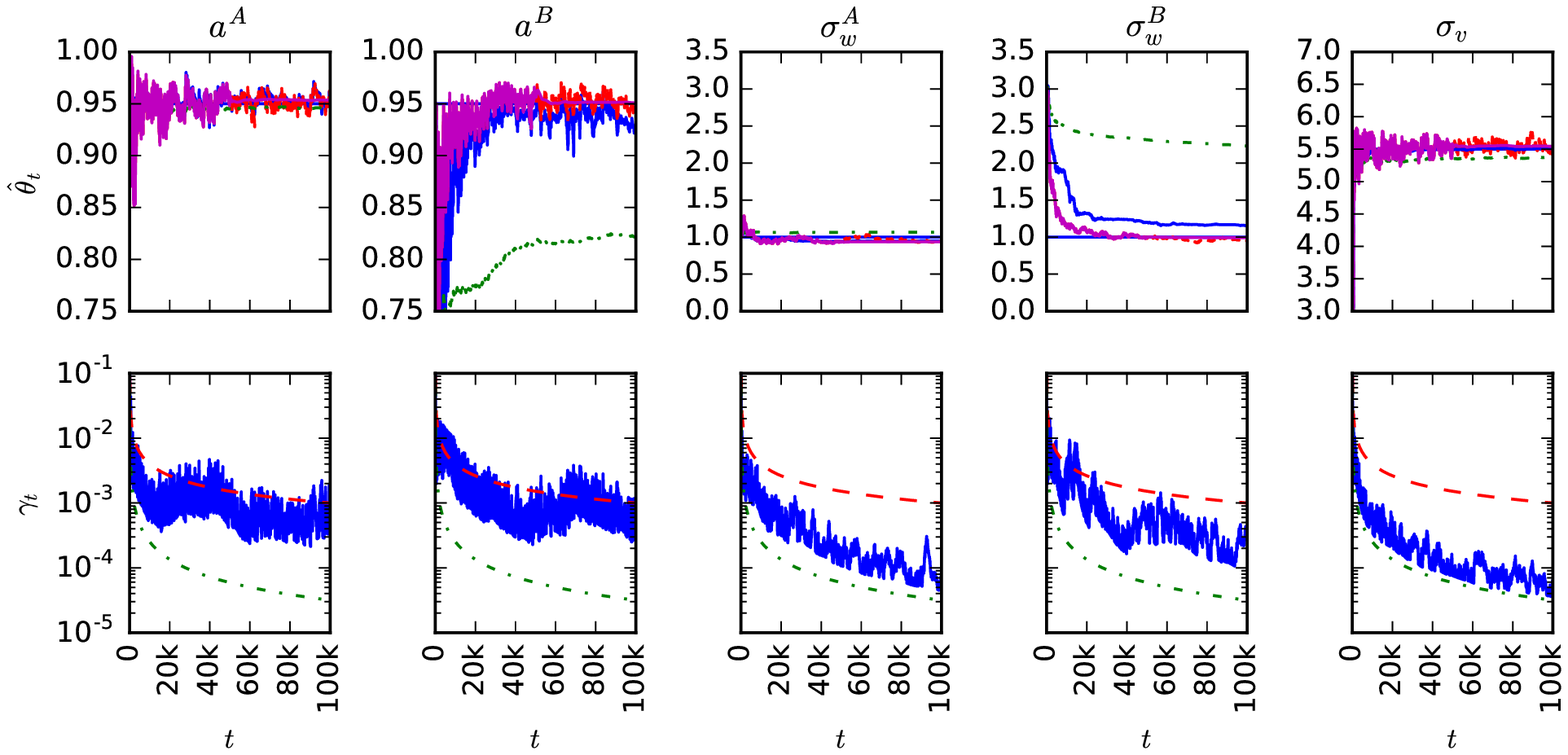}
\caption{Parameter-specific convergence in the 2-dimensional autoregressive model over 100,000 observations.
Each column displays information for a single parameter.
The top row shows the sequence of parameter estimates for three EM methods.
The bottom row shows the sequence of weights $\gamma_t$ for the three EM methods.
Blue solid line: IOEM; red dashed line: OEM with $c=0.6$; green dash-dot line: OEM with $c=0.9$;
magenta solid line: averaged OEM technique with a threshold $t_0=50,000$ }
\label{fig:sup4}
\end{figure*}

\begin{figure*}[!h]
\includegraphics[width=0.328\linewidth]{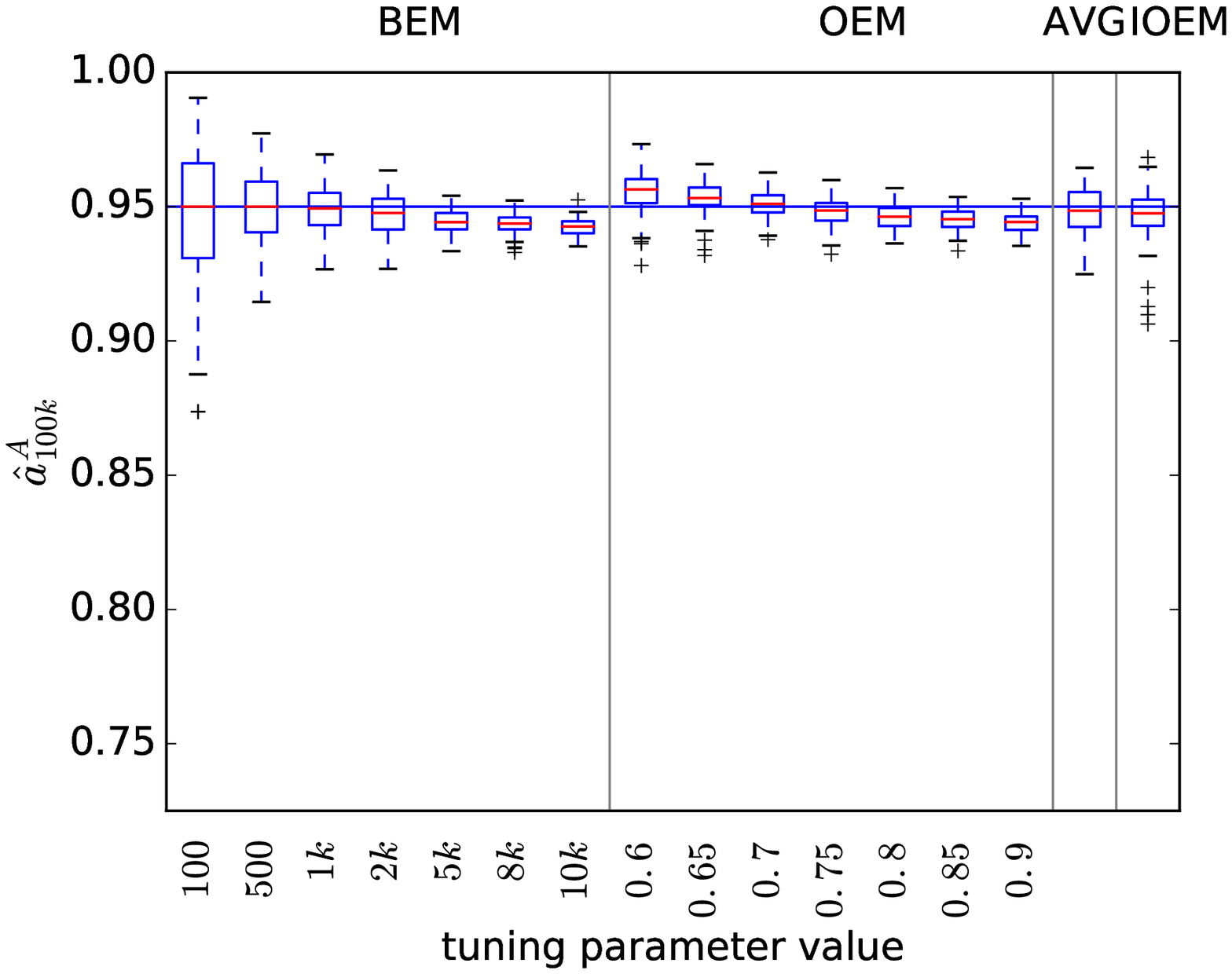}
\includegraphics[width=0.328\linewidth]{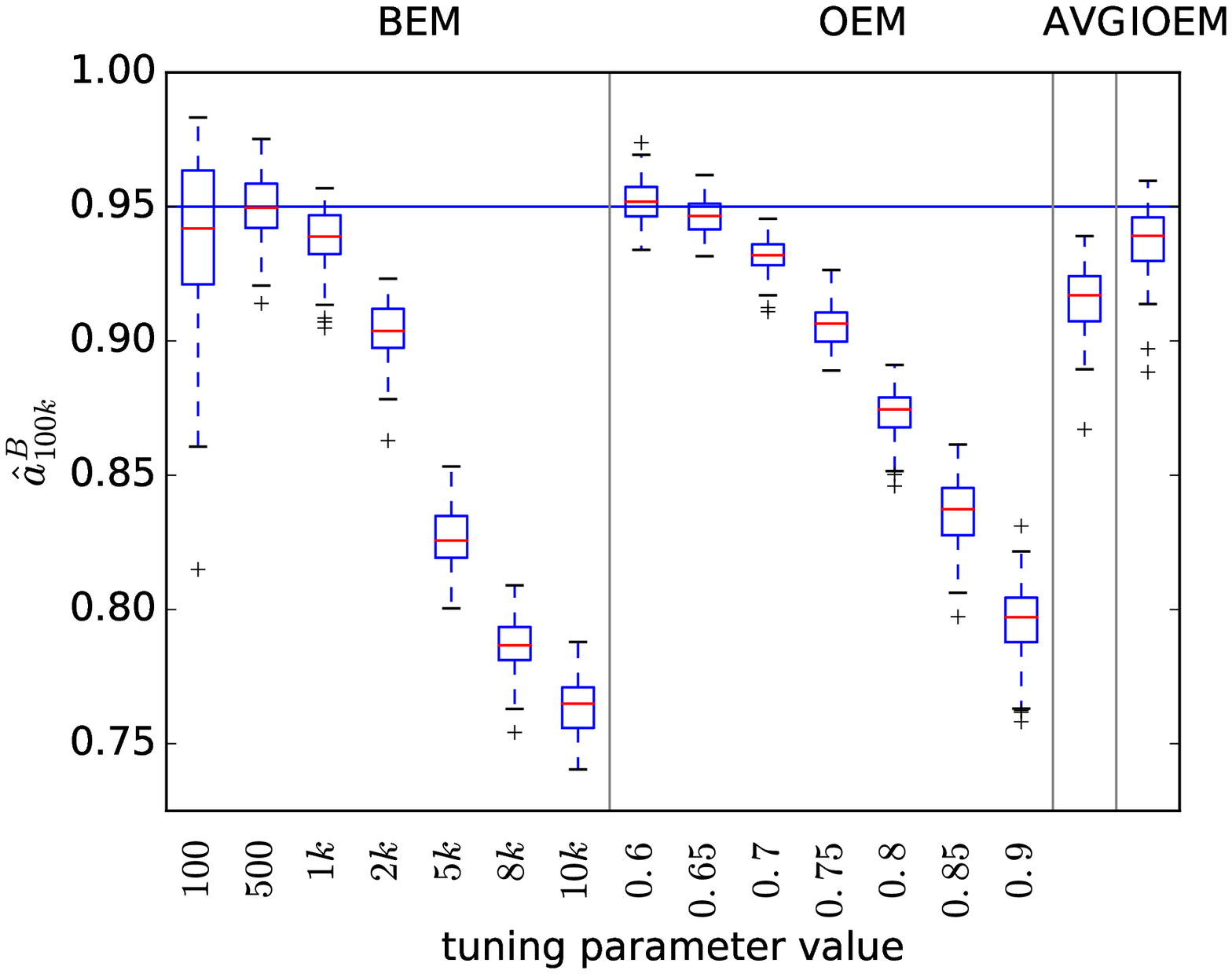}
\\
\includegraphics[width=0.328\linewidth]{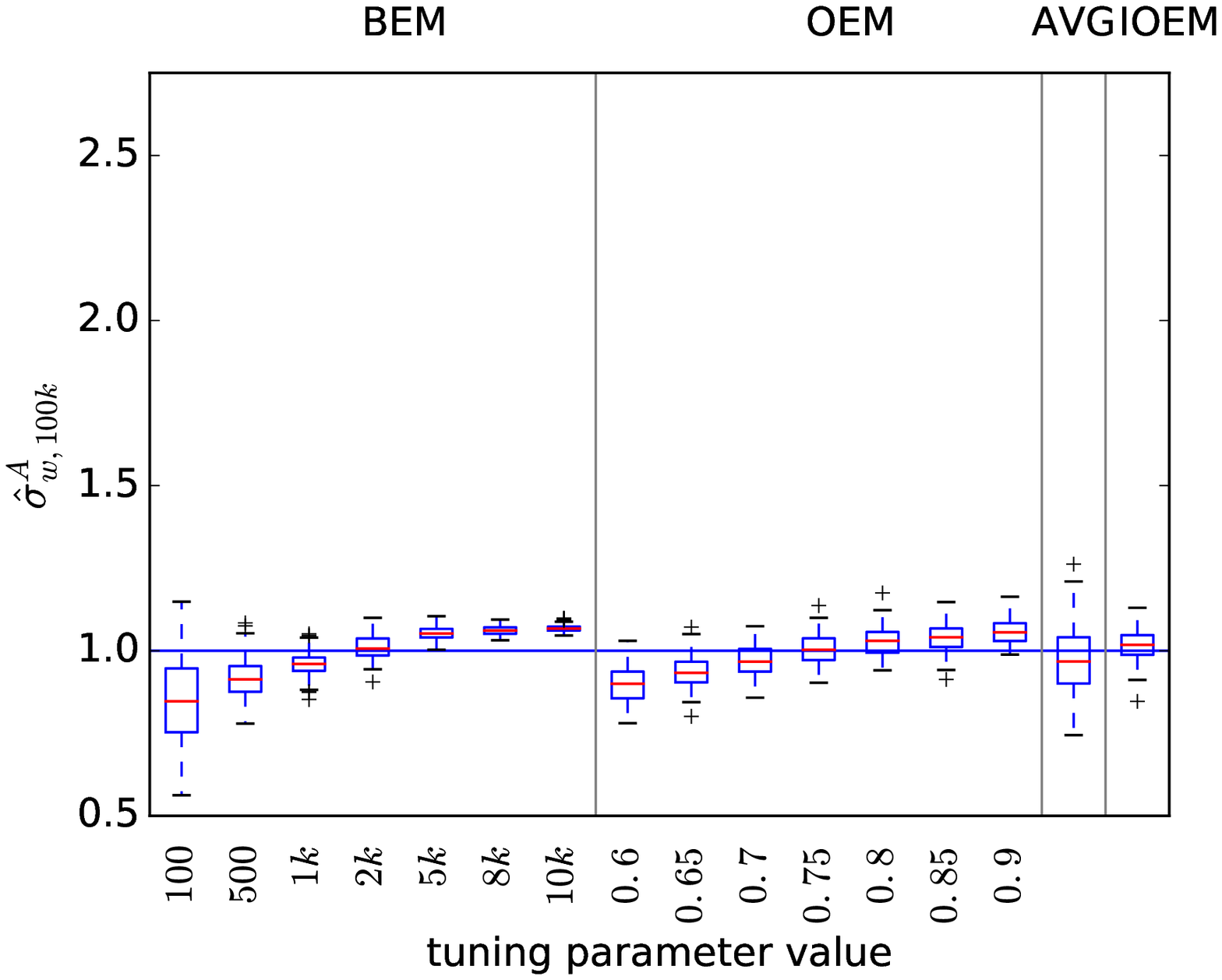}
\includegraphics[width=0.328\linewidth]{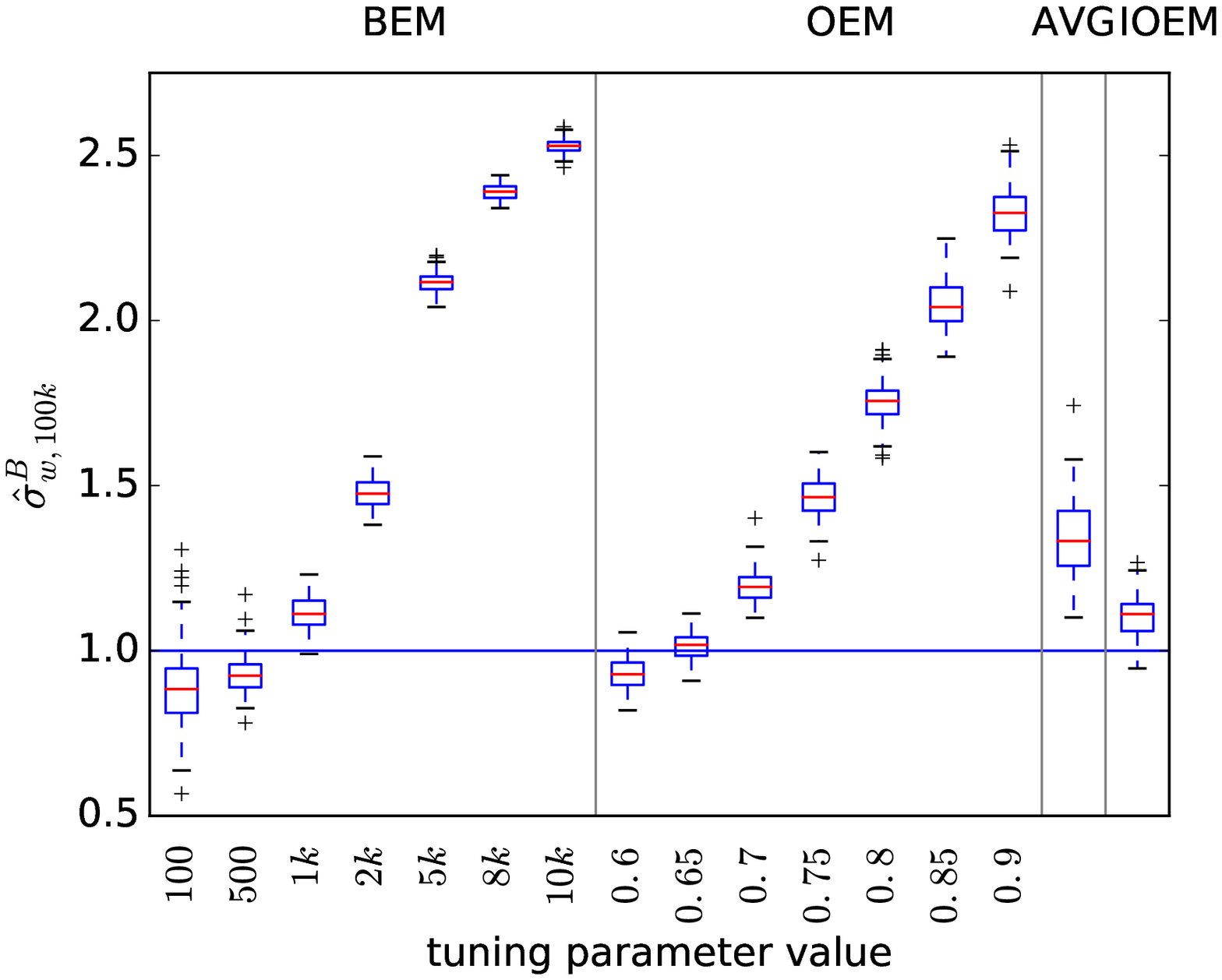}
\includegraphics[width=0.328\linewidth]{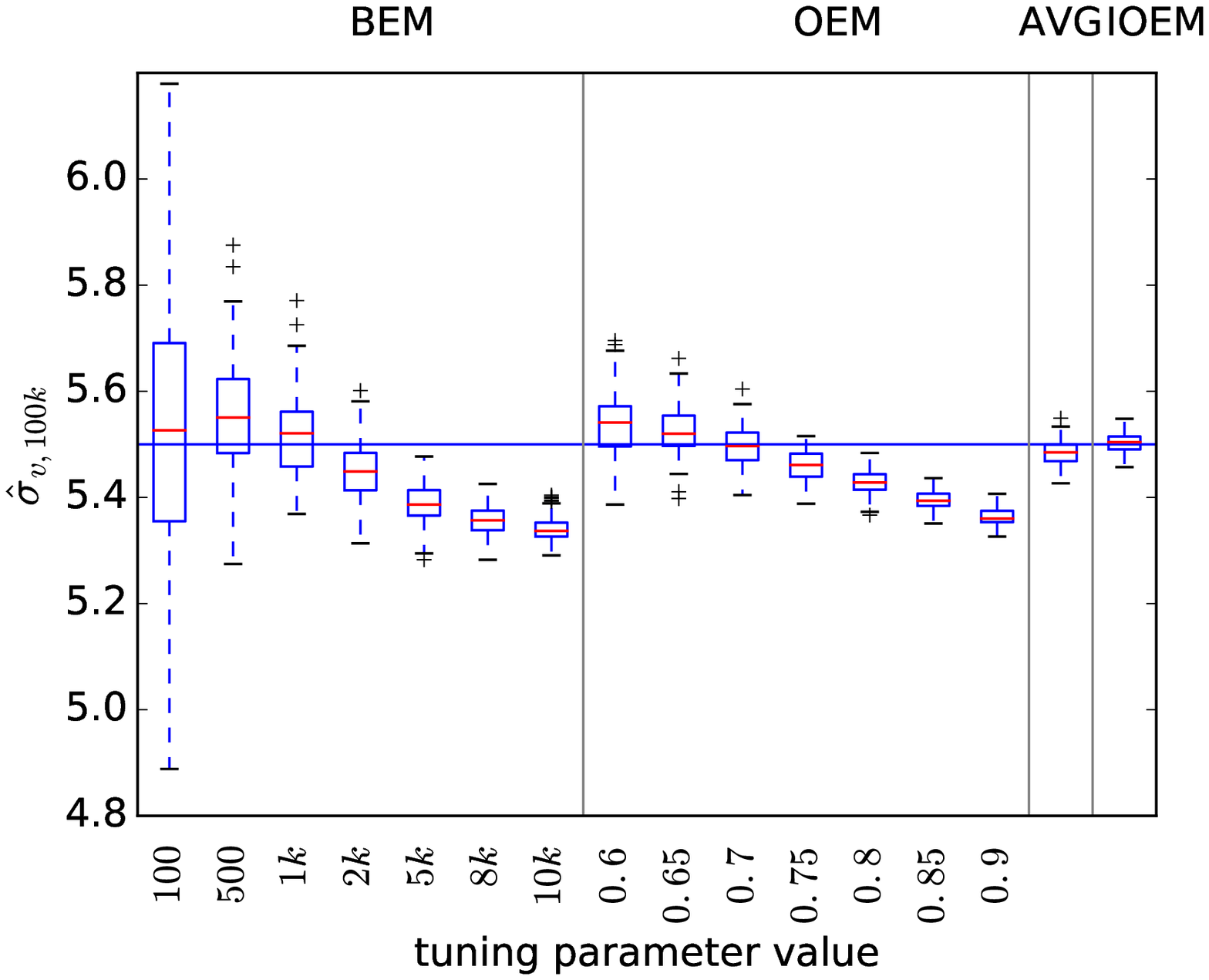}
\caption{Comparison of EM methods on 2-dimensional autoregressive model with true
parameters $a^A=0.95$, $\sigma^{A}_{w}=1$, $\sigma_{v}=5.5$, $a^B=0.95$, $\sigma^{B}_{w}=1$
and inital parameters $a^{A}_{0}=0.95$, $\sigma^{A}_{w,0}=1$, $\sigma_{v,0}=3$, $a^{B}_{0}=0.95$,
$\sigma^{B}_{w,0}=3$. Parameter estimates at $t=100,000$ are plotted for 100 replicates, $N=100$}
\label{fig:sup5}
\end{figure*}

\begin{figure*}[!h]
\includegraphics[width=.99\linewidth]{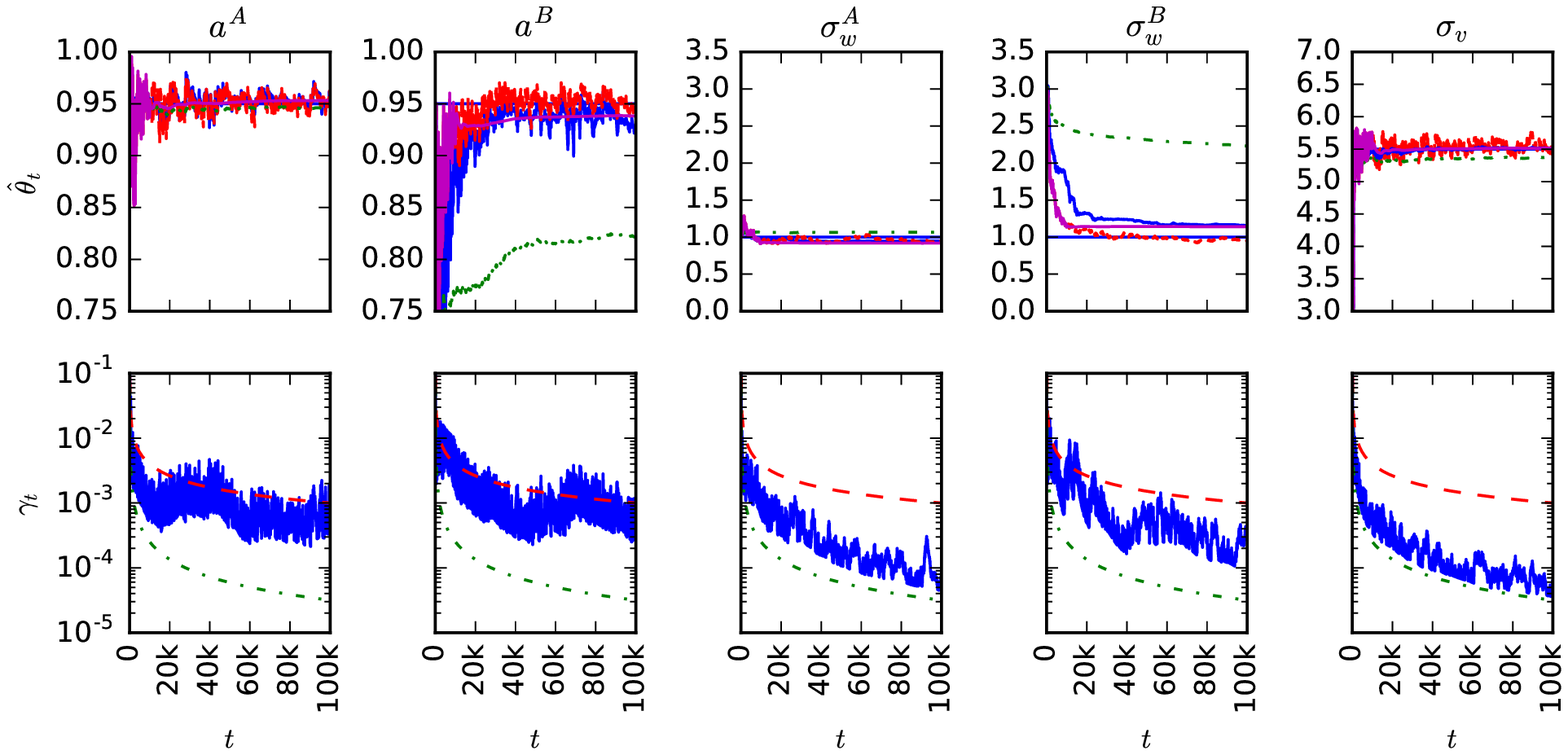}
\caption{Parameter-specific convergence in the 2-dimensional autoregressive model over 100,000 observations.
Each column displays information for a single parameter.
The top row shows the sequence of parameter estimates for four EM methods.
The bottom row shows the sequence of weights $\gamma_t$ for the three EM methods.
Blue solid line: IOEM; red dashed line: OEM with $c=0.6$; green dash-dot line: OEM with $c=0.9$;
magenta solid line: averaged OEM technique with a threshold $t_0=10,000$ }
\label{fig:sup6}
\end{figure*}

\begin{figure*}[!h]
\includegraphics[width=0.328\linewidth]{figures/sv_phi_100k_final.eps}
\includegraphics[width=0.328\linewidth]{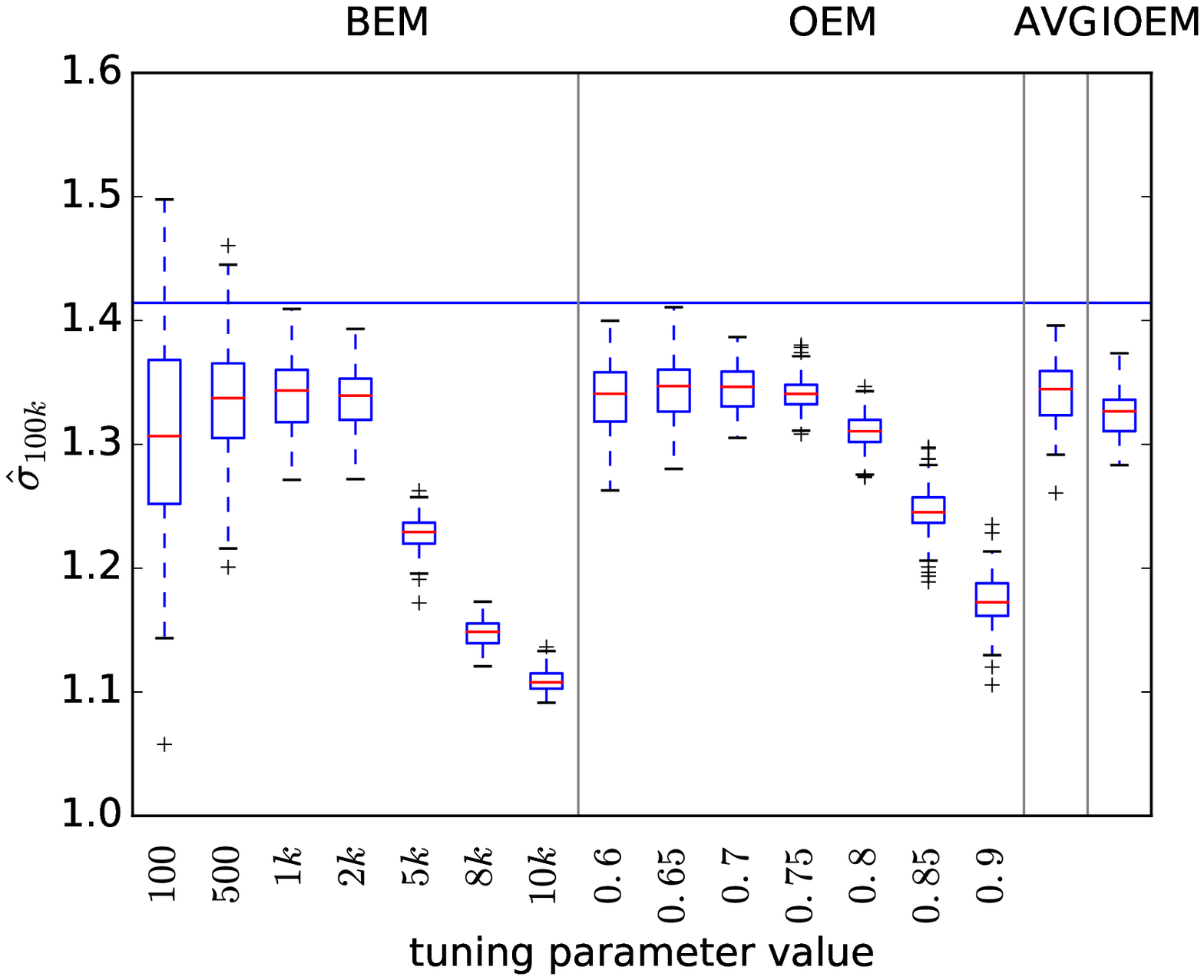}
\includegraphics[width=0.328\linewidth]{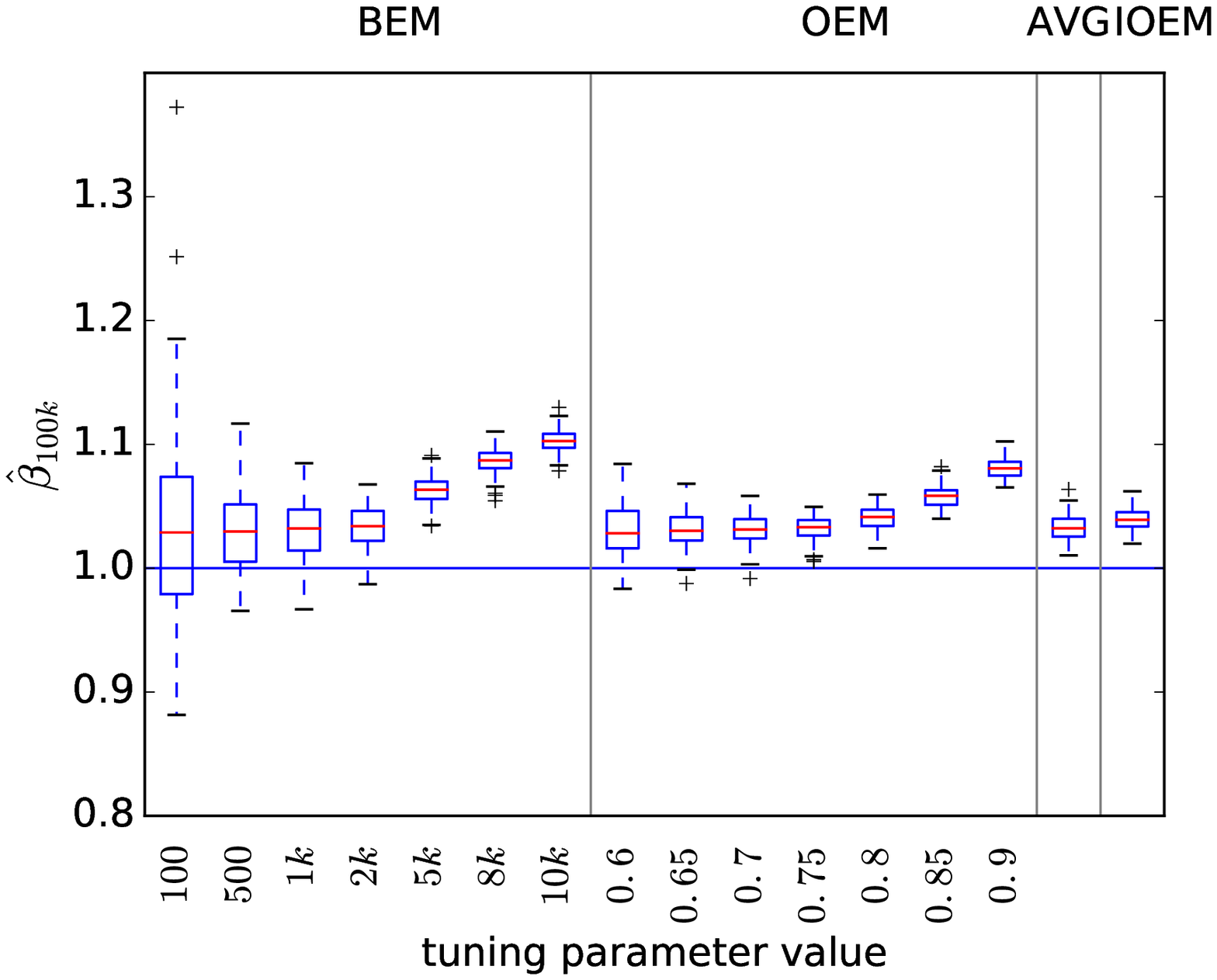}
\caption{Comparison of EM methods on stochastic volatility model
with unknown true parameters $\phi=0.1$, $\sigma=\sqrt{2}$, $\beta=1$  and inital parameters
$\phi_{0}=0.5$, $\sigma_{0}=1$, $\beta_{0}=\sqrt{2}$. Parameter estimates at $t=100,000$
are plotted for 100 replicates, $N=100$}
\label{fig:sup2}
\end{figure*}

\end{subsection}


\end{document}